\documentclass[aps,floats,twoside,fancyhdr,superscriptaddress,nofootinbib,showpacs]{revtex4}%
\usepackage{latexsym}
\usepackage{amsmath}
\usepackage{amssymb}
\usepackage{amsfonts}
\usepackage{graphicx}
\usepackage{dcolumn}
\usepackage[colorlinks]{hyperref}%

\newtheorem{theorem}{Theorem}

\newtheorem{remark}[theorem]{Remark}

\begin{document}
\preprint{gr-qc/0707.????}
\title[Bianchi I]{About Bianchi I with time varying constants.}
\author{Jos\'{e} Antonio Belinch\'{o}n}
\affiliation{Dept. of Physics. ETS Architecture. UPM Av. Juan de Herrera 4. Madrid 28040. Espa\~{n}a.}
\email{abelcal@ciccp.es}
\date{July 20, 2007 }

\begin{abstract}
In this paper we study how to attack through different techniques a perfect
fluid Bianchi I model with variable $G,$and $\Lambda.$ These tactics are: Lie
groups method (LM), imposing a particular symmetry, self-similarity (SS),
matter collineations (MC). and kinematical self-similarity (KSS). We compare
both tactics since they are quite similar (symmetry principles). We arrive to
the conclusion that the LM is too restrictive and brings us to get only the
flat FRW solution with $G=const.$ and $\Lambda=0.$ The SS, MC and KSS
approaches bring us to obtain the following solution: $G$ is a decreasing time
function and $\Lambda\thickapprox t^{-2}$, with $\Lambda<0, $ while the
exponents of the scale factor must satisfy the conditions $\sum_{i=1}%
^{3}\alpha_{i}=1$ and $\sum_{i=1}^{3}\alpha_{i}^{2}<1,$ $\forall\omega
\in\left(  -1,1\right)  ,$ relaxing in this way the Kasner conditions.

\end{abstract}

\maketitle
\tableofcontents


\vspace{.4cm}

\section{Introduction.}

Ever since Dirac first considered the possibility of a $G$ variable (see
\cite{1}), there have been numerous modifications of general relativity to
allow for a variable $G,$ nevertheless these theories have not gained wide
acceptance. However, recently (see \cite{2}-\cite{14}) a modification has been
proposed treating $G$ and $\Lambda$ as non-constants coupling scalars. So it
is considered $G$ and $\Lambda$ as coupling scalars within the Einstein
equations, $R_{ij}-\frac{1}{2}g_{ij}=GT_{ij}-\Lambda g_{ij},$ \ while the
other symbols have their usual meaning and hence the principle of equivalence
then demands that only $g_{ij}$ and not $G$ and $\Lambda$ must enter the
equation of motion of particles and photons. in this way the usual
conservation law, $divT=0,$ holds. Taking the divergence of the Einstein
equations and using the Bianchi identities we obtain the an equation that
controls the variation of $G$ and $\Lambda.$ These are the modified field
equations that allow to take into account a variable $G$ and $\Lambda.$
Nevertheless this approach has some drawbacks, for example, it cannot derived
from a Hamiltonian, although there are several advantages in the approach.

There are many publications devoted to study the variation of $G$ and
$\Lambda$ in the framework of flat FRW symmetries (see for example
\cite{2}-\cite{14}) and all this works have been extended to more complicated
geometries, like for example Bianchi I models, which represent the simplest
generalization of the flat FRW models (see for example \cite{Bes1}-\cite{SPS}.
in the context of perfect fluids and \cite{Arbab}-\cite{Saha1} in the context
of viscous fluids). Bianchi I models are important in the study of anisotropies.

But in our opinion, the problem arises when one try to solve the resulting
field equations (FE). It seems that it is unavoidable to make simplifying
hypotheses, or to impose ad hoc some particular behaviour for some of the
quantities of the model, in order to obtain a exact solution to the FE \ Such
simplifying hypothesis are made for mathematical reason (in order to reduce
the number of unknowns) although are justified form the physical point of
view. Usually such assumptions or simplifying hypothesis follow a power law,
for example, the quantity $X$ follows a power law \ i.e. $X=X_{0}t^{\alpha},$
where $X_{0}$ is an appropriate dimensional constant, $\ t$ is the cosmic time
(for example) and $\alpha\in\mathbb{R}$ (usually $\alpha\in\mathbb{Q},$ but
this other question), and depending on the nature of the quantity $X,$
$\alpha$ will be positive or negative. Actually we think that although all
these simplifying hypothesis are correct or at least bring us to obtain
correct results, it is not necessary to do that, since they may be deduced
from symmetry principles in such a way that one may justify (deduce) them from
a correct mathematical principle, and usually all these approaches have
physical meaning.

Therefore the main goal of this paper is to develop and compare some well
known tactics (approaches) in order to study and find exact solutions for a
perfect fluid Bianchi I models with variable $G$ and $\Lambda,$ but trying to
make the lowest number of assumptions or neither. We will try to show that
with these approaches all the usual simplifying hypotheses may be deduced from
a correct mathematical principle and how the useful are each tactic, i.e. to
show the advantages and disadvantages of each approach. We have started
studying this class of models because, as we have mentioned above, there are
many well known exact solutions so we will be able to compare the useful of
our approach. In a forthcoming paper we extend this study to a very
complicated model (from the mathematical point of view) which is a perfect
fluid Bianchi I within the framework of a variable speed of light (VSL) where
we have to study, for example a third order ODE with four unknowns, and we
need to be sure that theses tactics, the exposed ones here, work well. Hence
in this paper we are going to study a Bianchi I model with variable $G$ and
$\Lambda$ through the Lie group method (LM), studying the symmetries of the
resulting ODE, and through the self-similarity (SS), matter collineations (MC)
as well as kinematical self-similarity (KSS) hypothesis.

The paper is divided in the following sections: In section two we outline the
main ingredients of the model as well as the FE (under the condition
$divT=0).$ In order to apply the (LM) we need to deduce a ODE. For this
purpose we have followed model proposed by Kalligas, Wesson and Everitt,
\cite{We}, but taking into account some little differences. In section three
are calculated all the curvature tensors, Weyl etc... as well as their
invariants, i.e Kretschmann scalars etc.... In section four it is studied
through the Lie group tactic a third order differential equation with two
unknowns. \ We seek \ the possible forms that may take $G$ in other to make
integrable the ODE. In this way we find that there are three possibilities,
but the question here, is that all the studied solutions are unphysical (in
the sense that the shear vanish i.e. $\sigma=0$) or trivial. Actually this is
a very surprising result, since the third solution is very similar to the
obtained one in reference \cite{We}, but if one follows all the calculations
until the end, then arrive to the conclusion that such solution is the flat
FRW one with $G$ constant and with a vanishing $\Lambda$. Since in the case of
a $c-$var the third order ODE is quite complicate, in appendix A, we study the
resulting second order ODE, which is simpler than the third one but has the
drawback of having three unknowns. In this case is really difficult to arrive
to some conclusions since the obtained solutions depend of many integrating
constants although the solutions have or follow the same behavior, in order of
magnitude, as in the previous case.

In section five, we study the model under the self-similarity hypothesis. in
this case, the obtained solution are similar (in order of magnitude) to the
obtained one in the above section (LM with scaling symmetries). Nevertheless
in this case we are able to obtain a solution with shear non-vanishing i.e.
$\sigma\neq0,$ where $G$ is a decreasing time function and $\Lambda
\thickapprox t^{-2}$, with $\Lambda<0,$ while the exponents of the scale
factor must satisfy the conditions $\sum_{i=1}^{3}\alpha_{i}=1$ and
$\sum_{i=1}^{3}\alpha_{i}^{2}<1,$ $\forall\omega\in\left(  -1,1\right)  ,$
relaxing in this way the Kasner conditions..Since the model is SS, then, in
section six, we study the model studying the matter collineations (MC). In
this occasion we need to reformulate or reinterpret the MC equations in order
to get information on the behavior of $G$ and $\Lambda$, arriving to the same
conclusions as in the above section. In the last section we reproduce the same
tactic but this time under the KSS hypothesis, in this occasion we get a
non-singular solution and with the same behavior for the main quantities as
the obtained one in the above sections. We end this section discussing the
Kasner like solutions. We end with a brief conclusions.

We have add another appendix, appendix B, where we review a set of solutions
for Bianchi I models with time varying constants but under the hypothesis
$divT\neq0,$ as well as, the standard one, which is the model where $G=const.
$ and $\Lambda=0.$ We arrive to similar conclusions as the obtained ones in
the paper.

\section{The Model\label{Model}.}

Throughout the paper $M$ will denote the usual smooth (connected, Hausdorff,
4-dimensional) spacetime manifold with smooth Lorentz metric $g$ of signature
$(-,+,+,+)$. Thus $M$ is paracompact. A comma, semi-colon and the symbol
$\mathcal{L}$ denote the usual partial, covariant and Lie derivative,
respectively, the covariant derivative being with respect to the Levi-Civita
connection on $M$ derived from $g$. The associated Ricci and stress-energy
tensors will be denoted in component form by $R_{ij}(\equiv R^{c}{}_{jcd})$
and $T_{ij}$ respectively. A diagonal Bianchi I space-time is a spatially
homogeneous space-time which admits an abelian group of isometries $G_{3}$,
acting on spacelike hypersurfaces, generated by the spacelike KVs
$\mathbf{\xi}_{1}=\partial_{x},\mathbf{\xi}_{2}=\partial_{y},\mathbf{\xi}%
_{3}=\partial_{z}$. In synchronous co-ordinates the metric is:%
\begin{equation}
ds^{2}=-dt^{2}+A_{\mu}^{2}(t)(dx^{\mu})^{2}\label{sx1.2}%
\end{equation}
where the metric functions $A_{1}(t),A_{2}(t),A_{3}(t)$ are functions of the
time co-ordinate only (Greek indices take the space values $1,2,3$ and Latin
indices the space-time values $0,1,2,3$). In this paper we are interested only
in \emph{proper diagonal} Bianchi I space-times (which in the following will
be referred for convenience simply as Bianchi I\ space-times), hence all
metric functions are assumed to be different and the dimension of the group of
isometries acting on the spacelike hypersurfaces is three. Therefore we
consider the Bianchi type I metric as
\begin{equation}
ds^{2}=-c^{2}dt^{2}+X^{2}(t)dx^{2}+Y^{2}(t)dy^{2}+Z^{2}(t)dz^{2},\label{eq1}%
\end{equation}
see for example (\cite{em}-\cite{Raycha}).

For a perfect fluid with energy-momentum tensor:
\begin{equation}
T_{ij}=\left(  \rho+p\right)  u_{i}u_{j}+pg_{ij},\label{eq0}%
\end{equation}
where we are assuming an equation of state $p=\omega\rho,\left(
\omega=const.\right)  $. Note that here we have preferred to assume this
equation of state but as we will show in the following sections this equation
may be deduced from the symmetries principles as for example the self-similar
one. The $4-$velocity is defined as follows%
\begin{equation}
u=\left(  \frac{1}{c},0,0,0\right)  ,\qquad u_{i}u^{i}=-1.\label{4-vel}%
\end{equation}

The time derivatives of $G$ and $\Lambda$ are related by the Bianchi
identities
\begin{equation}
\left(  R_{ij}-\frac{1}{2}Rg_{ij}\right)  ^{;j}=\left(  \frac{8\pi G}{c^{4}%
}T_{ij}-\Lambda g_{ij}\right)  ^{;j},\label{eq8}%
\end{equation}
in this case this equation reads:%
\begin{equation}
\frac{8\pi G}{c^{4}}\left[  \dot{\rho}+\rho\left(  1+\omega\right)  \left(
\frac{\dot{X}}{X}+\frac{\dot{Y}}{Y}+\frac{\dot{Z}}{Z}\right)  \right]
=-\dot{\Lambda}-\frac{8\pi}{c^{4}}\dot{G}\rho,\label{laura3}%
\end{equation}
in our case we obtain (taking into account the additional condition $\left(
divT=0\right)  ):$%
\begin{align}
\dot{\rho}+\rho\left(  1+\omega\right)  \left(  \frac{\dot{X}}{X}+\frac
{\dot{Y}}{Y}+\frac{\dot{Z}}{Z}\right)   &  =0,\\
\dot{\Lambda} &  =-\frac{8\pi}{c^{4}}\dot{G}\rho,
\end{align}
in the appendix \ref{ApB} we will study how to obtain different cosmological
models, under the self-similarity hypothesis, with one or several time varying
\textquotedblleft constants\textquotedblright\ but without the assumed
condition $\operatorname{div}T=0.$

Therefore the resulting field equations are:
\begin{align}
\frac{\dot{X}}{X}\frac{\dot{Y}}{Y}+\frac{\dot{X}}{X}\frac{\dot{Z}}{Z}%
+\frac{\dot{Z}}{Z}\frac{\dot{Y}}{Y}  &  =\frac{8\pi G}{c^{2}}\rho+\Lambda
c^{2},\label{eq3}\\
\frac{\ddot{Y}}{Y}+\frac{\ddot{Z}}{Z}+\frac{\dot{Z}}{Z}\frac{\dot{Y}}{Y}  &
=-\frac{8\pi G}{c^{2}}\omega\rho+\Lambda c^{2},\label{eq4}\\
\frac{\ddot{X}}{X}+\frac{\ddot{Z}}{Z}+\frac{\dot{X}}{X}\frac{\dot{Z}}{Z}  &
=-\frac{8\pi G}{c^{2}}\omega\rho+\Lambda c^{2},\label{eq5}\\
\frac{\ddot{X}}{X}+\frac{\ddot{Y}}{Y}+\frac{\dot{X}}{X}\frac{\dot{Y}}{Y}  &
=-\frac{8\pi G}{c^{2}}\omega\rho+\Lambda c^{2},\label{eq6}\\
\dot{\rho}+\rho\left(  1+\omega\right)  \left(  \frac{\dot{X}}{X}+\frac
{\dot{Y}}{Y}+\frac{\dot{Z}}{Z}\right)   &  =0,\label{eq7}\\
\dot{\Lambda}  &  =-\frac{8\pi}{c^{4}}\dot{G}\rho.\label{eq9}%
\end{align}

Now, if we define
\begin{equation}
H=\left(  \frac{\dot{X}}{X}+\frac{\dot{Y}}{Y}+\frac{\dot{Z}}{Z}\right)
=3\frac{\dot{R}}{R}\text{ \ and \ \ }R^{3}=XYZ,\qquad q=\frac{d}{dt}\left(
\frac{1}{H}\right)  -1,\label{eq14}%
\end{equation}
it is regained eq.(\ref{eq7}) as the usual conservation equation i.e.
\begin{equation}
\dot{\rho}+\rho\left(  1+\omega\right)  H=0.
\end{equation}

Since we have defined the 4-velocity by eq. (\ref{4-vel}) then the expansion
$\theta$ is defined as follows:
\begin{equation}
\theta:=u_{;i}^{i},\text{ \ \ \ \ \ \ \ \ }\theta=\frac{1}{c}\left(
\frac{\dot{X}}{X}+\frac{\dot{Y}}{Y}+\frac{\dot{Z}}{Z}\right)  =\frac{1}{c}H,
\end{equation}
and therefore the acceleration is:%
\begin{equation}
a_{i}=u_{i;j}u^{j},
\end{equation}
in this case $a=0,$ while the shear is defined as follows:%
\begin{equation}
\sigma_{ij}=\frac{1}{2}\left(  u_{i;a}h_{j}^{a}+u_{j;a}h_{i}^{a}\right)
-\frac{1}{3}\theta h_{ij},
\end{equation}%
\begin{equation}
\sigma^{2}=\frac{1}{2}\sigma_{ij}\sigma^{ij},\text{\ \ \ \ \ \ }\sigma
^{2}=\frac{1}{3c^{2}}\left(  \left(  \frac{\dot{X}}{X}\right)  ^{2}+\left(
\frac{\dot{Y}}{Y}\right)  ^{2}+\left(  \frac{\dot{Z}}{Z}\right)  ^{2}%
-\frac{\dot{X}}{X}\frac{\dot{Y}}{Y}-\frac{\dot{X}}{X}\frac{\dot{Z}}{Z}%
-\frac{\dot{Y}}{Y}\frac{\dot{Z}}{Z}\right)  ,\label{defshear}%
\end{equation}
or equivalently%
\begin{equation}
\sigma^{2}=\frac{1}{6c^{2}}\left(  \left(  \frac{\dot{X}}{X}-\frac{\dot{Y}}%
{Y}\right)  ^{2}+\left(  \frac{\dot{X}}{X}-\frac{\dot{Z}}{Z}\right)
^{2}+\left(  \frac{\dot{Y}}{Y}-\frac{\dot{Z}}{Z}\right)  ^{2}\right)
.\label{defshear2}%
\end{equation}

\subsection{The crucial equations.\label{CE}}

In this section we would like to obtain a ODE which allows us to study all the
field equations through the Lie method. For this purpose we are following
closely the paper by Kalligas et al (see \cite{We}). From eqs. (\ref{eq4}%
-\ref{eq6}) and taking into account eq. (\ref{eq3}), we get:
\begin{equation}
\frac{\ddot{X}}{X}+\frac{\ddot{Y}}{Y}+\frac{\ddot{Z}}{Z}=-\frac{4\pi}{c^{2}%
}\left(  1+3\omega\right)  G\rho+\Lambda c^{2}.\label{w1}%
\end{equation}

Now, taking into account eq. (\ref{eq7}), squaring it we get%
\begin{equation}
\left(  \frac{\dot{\rho}}{\rho}\right)  ^{2}=\left(  1+\omega\right)
^{2}H^{2},
\end{equation}
since%
\begin{equation}
H^{2}=\left(  \left(  \frac{\dot{X}}{X}\right)  ^{2}+\left(  \frac{\dot{Y}}%
{Y}\right)  ^{2}+\left(  \frac{\dot{Z}}{Z}\right)  ^{2}+2\left(  \frac{\dot
{X}}{X}\frac{\dot{Y}}{Y}+\frac{\dot{X}}{X}\frac{\dot{Z}}{Z}+\frac{\dot{Z}}%
{Z}\frac{\dot{Y}}{Y}\right)  \right)  ,\label{eq-H}%
\end{equation}
and taking into account (\ref{eq3}) we get
\begin{equation}
\left(  \frac{\dot{\rho}}{\rho}\right)  ^{2}=\left(  1+\omega\right)
^{2}\left(  \left(  \frac{\dot{X}}{X}\right)  ^{2}+\left(  \frac{\dot{Y}}%
{Y}\right)  ^{2}+\left(  \frac{\dot{Z}}{Z}\right)  ^{2}+\frac{16\pi}{c^{2}%
}G\rho+2\Lambda c^{2}\right)  .\label{w2}%
\end{equation}

The time derivative $\frac{\dot{\rho}}{\rho}$ from eq. (\ref{eq7}) can now be
expressed in terms of $G,\Lambda$ and $\rho$ only by using eqs. (\ref{w1}) and
(\ref{w2}), it is obtained:
\begin{equation}
\left(  \frac{\dot{\rho}}{\rho}\right)  ^{2}\frac{2+\omega}{\left(
1+\omega\right)  ^{2}}-\frac{\ddot{\rho}}{\rho\left(  1+\omega\right)  }%
=\frac{12\pi}{c^{2}}G\rho\left(  1-\omega\right)  +3\Lambda c^{2}.\label{w3}%
\end{equation}

Now if we rewrite this equation in an appropriate way we get the following one
that we will also study through the Lie method (see appendix \ref{ApA}).. i.e.%
\begin{equation}
\ddot{\rho}=K_{1}\frac{\dot{\rho}^{2}}{\rho}+K_{2}G\rho^{2}-K_{3}\Lambda
\rho,\label{New-Lie}%
\end{equation}
where%
\begin{equation}
K_{1}=\frac{2+\omega}{\left(  1+\omega\right)  },\qquad K_{2}=\frac
{12\pi\left(  \omega^{2}-1\right)  }{c^{2}},\qquad K_{3}=3\left(
1+\omega\right)  c^{2}.
\end{equation}
as it is observed if $\omega=1,$ then $K_{2}=0,$ so eq.(\ref{New-Lie})
collapses to a very simple ODE with only two unknowns.

In order to reduce the number of unknowns, now, on differentiating eq.
(\ref{w3}) and taking into account eq. (\ref{eq9}) we arrive to the following
equation%
\begin{equation}
\dddot{\rho}=K_{1}\ddot{\rho}\frac{\dot{\rho}}{\rho}-K_{2}\frac{\dot{\rho}%
^{3}}{\rho^{2}}+K_{3}\dot{G}\rho^{2}-K_{4}G\dot{\rho}\rho,\label{defeq}%
\end{equation}
where%
\begin{equation}
K_{1}=\frac{5+3\omega}{1+\omega},\qquad K_{2}=\frac{4+2\omega}{1+\omega
},\qquad K_{3}=\frac{12\pi\left(  1+\omega\right)  ^{2}}{c^{2}},\qquad
K_{4}=\frac{12\pi\left(  1-\omega^{2}\right)  }{c^{2}},\label{monica}%
\end{equation}
we are supposing that $\omega\neq-1.$ This equation i.e. eq. (\ref{defeq})
will be studied exhaustively through the Lie method in section \ref{LM}.

As it is observed if $\omega=1$ (ultra-stiff matter) eq. (\ref{defeq})
collapses to%
\begin{equation}
\dddot{\rho}=4\ddot{\rho}\frac{\dot{\rho}}{\rho}-3\frac{\dot{\rho}^{3}}%
{\rho^{2}}+\frac{48\pi}{c^{2}}\dot{G}\rho^{2}.\label{case1}%
\end{equation}

The shear is calculated as follows. If we take into account the definition of
the shear i.e. eq. (\ref{defshear2}) as well as eqs. (\ref{eq3}),(\ref{eq7})
and (\ref{eq-H}) then we arrive to the following expression
\begin{equation}
\sigma^{2}=\frac{2}{6c^{2}}H^{2}-\left(  8\pi\frac{G}{c^{4}}\rho
+\Lambda\right)  =\frac{1}{3c^{2}\left(  1+\omega\right)  ^{2}}\left(
\frac{\dot{\rho}}{\rho}\right)  ^{2}-8\pi\frac{G}{c^{4}}\rho-\Lambda
\label{mshear}%
\end{equation}
note that $\omega\neq-1.$ We are only interested in solutions which verify
$\sigma\neq0.$

\begin{remark}
In reference (\cite{We}) the authors define the shear in a different way (see
eq. 14 of reference (\cite{We})). They define the anisotropy energy as%
\begin{equation}
8\pi G\sigma=6\left(  3\left(  \frac{R^{\prime}}{R}\right)  ^{2}-\left(
8\pi\frac{G}{c^{4}}\rho+\Lambda\right)  \right)  ,\label{kweshear}%
\end{equation}
which is similar to our definition of shear (compare with eq. \ref{mshear}).
\end{remark}

\section{Curvature behavior.\label{CB}}

The full contraction of the Riemann tensor (see for example \cite{Caminati}%
-\cite{Barrow})
\begin{equation}
I_{1}:=R_{ijkl}R^{ijkl},
\end{equation}%
\begin{equation}
I_{1}=\frac{4}{c^{4}}\left[  \left(  \frac{\ddot{X}}{X}\right)  ^{2}+\left(
\frac{\ddot{Y}}{Y}\right)  ^{2}+\left(  \frac{\ddot{Z}}{Z}\right)
^{2}+\left(  \frac{\dot{X}}{X}\frac{\dot{Y}}{Y}\right)  ^{2}+\left(
\frac{\dot{X}}{X}\frac{\dot{Z}}{Z}\right)  ^{2}+\left(  \frac{\dot{Y}}{Y}%
\frac{\dot{Z}}{Z}\right)  ^{2}\right]  ,
\end{equation}

The full contraction of the Ricci tensor%
\begin{equation}
I_{2}:=R_{ij}R^{ij},
\end{equation}%
\begin{align}
I_{2} &  =\frac{2}{c^{4}}\left[  \left(  \frac{\ddot{X}}{X}\right)
^{2}+\left(  \frac{\ddot{Y}}{Y}\right)  ^{2}+\left(  \frac{\ddot{Z}}%
{Z}\right)  ^{2}+\left(  \frac{\ddot{X}}{X}\frac{\ddot{Y}}{Y}\right)  +\left(
\frac{\ddot{X}}{X}\frac{\ddot{Z}}{Z}\right)  +\left(  \frac{\ddot{Y}}{Y}%
\frac{\ddot{Z}}{Z}\right)  +\left(  \frac{\ddot{X}}{X}\frac{\dot{X}}{X}%
\frac{\dot{Y}}{Y}\right)  +\right. \nonumber\\
&  +\left(  \frac{\ddot{X}}{X}\frac{\dot{X}}{X}\frac{\dot{Z}}{Z}\right)
+\left(  \frac{\dot{Y}}{Y}\frac{\dot{X}}{X}\right)  ^{2}+\left(  \frac{\dot
{X}}{X}\right)  ^{2}\frac{\dot{Y}}{Y}\frac{\dot{Z}}{Z}+\left(  \frac{\dot{Z}%
}{Z}\frac{\dot{X}}{X}\right)  ^{2}+\left(  \frac{\ddot{Y}}{Y}\frac{\dot{Y}}%
{Y}\frac{\dot{X}}{X}\right)  +\left(  \frac{\ddot{Y}}{Y}\frac{\dot{Y}}{Y}%
\frac{\dot{Z}}{Z}\right)  +\nonumber\\
&  \left.  +\frac{\dot{X}}{X}\left(  \frac{\dot{Y}}{Y}\right)  ^{2}\frac
{\dot{Z}}{Z}+\left(  \frac{\dot{Y}}{Y}\frac{\dot{Z}}{Z}\right)  ^{2}%
+\frac{\ddot{Z}}{Z}\frac{\dot{Z}}{Z}\frac{\dot{X}}{X}+\frac{\ddot{Z}}{Z}%
\frac{\dot{Z}}{Z}\frac{\dot{Y}}{Y}+\left(  \frac{\dot{Z}}{Z}\right)  ^{2}%
\frac{\dot{Y}}{Y}\frac{\dot{X}}{X}\right]  ,
\end{align}
and the scalar curvature $R$ is:%
\begin{equation}
R=\frac{2}{c^{2}}\left(  \frac{X^{\prime\prime}}{X}+\frac{Y^{\prime\prime}}%
{Y}+\frac{Z^{\prime\prime}}{Z}+\frac{X^{\prime}Y^{\prime}}{XY}+\frac
{Y^{\prime}Z^{\prime}}{YZ}+\frac{X^{\prime}Z^{\prime}}{XZ}\right)  .
\end{equation}

We calculate the Weyl tensor as well as the magnetic and the electric part of
the Weyl tensor.

The non-zero components of the Weyl tensor are:%
\begin{align}
C_{1212}  &  =\frac{X^{2}}{6}\left(  -2\frac{X^{\prime\prime}}{X}%
+\frac{X^{\prime}}{X}\left(  \frac{Y^{\prime}}{Y}+\frac{Z^{\prime}}{Z}\right)
+\frac{Y^{\prime\prime}}{Y}+\frac{Z^{\prime\prime}}{Z}-2\frac{Y^{\prime}}%
{Y}\frac{Z^{\prime}}{Z}\right)  ,\\
C_{1313}  &  =\frac{Y^{2}}{6}\left(  -2\frac{Y^{\prime\prime}}{Y}%
+\frac{Y^{\prime}}{Y}\left(  \frac{X^{\prime}}{X}+\frac{Z^{\prime}}{Z}\right)
+\frac{X^{\prime\prime}}{X}+\frac{Z^{\prime\prime}}{Z}-2\frac{X^{\prime}}%
{X}\frac{Z^{\prime}}{Z}\right)  ,\\
C_{1414}  &  =\frac{Z^{2}}{6}\left(  -2\frac{Z^{\prime\prime}}{Z}%
+\frac{Z^{\prime}}{Z}\left(  \frac{X^{\prime}}{X}+\frac{Y^{\prime}}{Y}\right)
+\frac{X^{\prime\prime}}{X}+\frac{Y^{\prime\prime}}{Y}-2\frac{X^{\prime}}%
{X}\frac{Y^{\prime}}{Y}\right)  ,\\
C_{2323}  &  =-\frac{X^{2}Y^{2}}{6c^{2}}\left(  -2\frac{Z^{\prime\prime}}%
{Z}+\frac{Z^{\prime}}{Z}\left(  \frac{X^{\prime}}{X}+\frac{Y^{\prime}}%
{Y}\right)  +\frac{X^{\prime\prime}}{X}+\frac{Y^{\prime\prime}}{Y}%
-2\frac{X^{\prime}}{X}\frac{Y^{\prime}}{Y}\right)  ,\\
C_{2424}  &  =-\frac{X^{2}Z^{2}}{6c^{2}}\left(  -2\frac{Y^{\prime\prime}}%
{Y}+\frac{Y^{\prime}}{Y}\left(  \frac{X^{\prime}}{X}+\frac{Z^{\prime}}%
{Z}\right)  +\frac{X^{\prime\prime}}{X}+\frac{Z^{\prime\prime}}{Z}%
-2\frac{X^{\prime}}{X}\frac{Z^{\prime}}{Z}\right)  ,\\
C_{3434}  &  =-\frac{Y^{2}Z^{2}}{6c^{2}}\left(  -2\frac{X^{\prime\prime}}%
{X}+\frac{X^{\prime}}{X}\left(  \frac{Y^{\prime}}{Y}+\frac{Z^{\prime}}%
{Z}\right)  +\frac{Y^{\prime\prime}}{Y}+\frac{Z^{\prime\prime}}{Z}%
-2\frac{Y^{\prime}}{Y}\frac{Z^{\prime}}{Z}\right)  ,
\end{align}
where $X^{\prime}:=\dot{X}.$

The non-zero components of the electric part of the Weyl tensor are:%
\begin{align}
E_{22}  &  =\frac{X^{2}}{6c^{2}}\left(  -2\frac{X^{\prime\prime}}{X}%
+\frac{X^{\prime}}{X}\left(  \frac{Y^{\prime}}{Y}+\frac{Z^{\prime}}{Z}\right)
+\left(  \frac{Y^{\prime\prime}}{Y}+\frac{Z^{\prime\prime}}{Z}-2\frac
{Y^{\prime}}{Y}\frac{Z^{\prime}}{Z}\right)  \right)  ,\\
E_{33}  &  =\frac{Y^{2}}{6c^{2}}\left(  -2\frac{Y^{\prime\prime}}{Y}%
+\frac{Y^{\prime}}{Y}\left(  \frac{X^{\prime}}{X}+\frac{Z^{\prime}}{Z}\right)
+\left(  \frac{X^{\prime\prime}}{X}+\frac{Z^{\prime\prime}}{Z}-2\frac
{X^{\prime}}{X}\frac{Z^{\prime}}{Z}\right)  \right)  ,\\
E_{44}  &  =\frac{Z^{2}}{6c^{2}}\left(  -2\frac{Z^{\prime\prime}}{Z}%
+\frac{Z^{\prime}}{Z}\left(  \frac{X^{\prime}}{X}+\frac{Y^{\prime}}{Y}\right)
+\left(  \frac{X^{\prime\prime}}{X}+\frac{Y^{\prime\prime}}{Y}-2\frac
{X^{\prime}}{X}\frac{Y^{\prime}}{Y}\right)  \right)  .
\end{align}

The magnetic part of the Weyl tensor vanish\textbf{\ }%
\begin{equation}
H_{ij}=0.
\end{equation}

The Weyl scalar is defined as:%
\begin{equation}
I_{3}:=C^{abcd}C_{abcd},
\end{equation}%
\begin{align}
I_{3} &  =\frac{4}{3c^{4}}\left[  -\frac{X^{\prime\prime}}{X}\frac
{Y^{\prime\prime}}{Y}-\frac{X^{\prime\prime}}{X}\frac{Z^{\prime\prime}}%
{Z}-\frac{Y^{\prime\prime}}{Y}\frac{Z^{\prime\prime}}{Z}+\left(
\frac{X^{\prime\prime}}{X}\right)  ^{2}+\left(  \frac{Y^{\prime\prime}}%
{Y}\right)  ^{2}+\left(  \frac{Z^{\prime\prime}}{Z}\right)  ^{2}\right.
\nonumber\\
&  +\frac{X^{\prime}}{X}\frac{Y^{\prime}}{Y}\left(  2\frac{Z^{\prime\prime}%
}{Z}-\left(  \frac{Z^{\prime}}{Z}\right)  ^{2}-\frac{X^{\prime\prime}}%
{X}-\frac{Y^{\prime\prime}}{Y}\right)  +\frac{Y^{\prime}}{Y}\frac{Z^{\prime}%
}{Z}\left(  2\frac{X^{\prime\prime}}{X}-\left(  \frac{X^{\prime}}{X}\right)
^{2}-\frac{Z^{\prime\prime}}{Z}-\frac{Y^{\prime\prime}}{Y}\right) \nonumber\\
&  \left.  +\frac{X^{\prime}}{X}\frac{Z^{\prime}}{Z}\left(  2\frac
{Y^{\prime\prime}}{Y}-\left(  \frac{Y^{\prime}}{Y}\right)  ^{2}-\frac
{X^{\prime\prime}}{X}-\frac{Z^{\prime\prime}}{Z}\right)  +\left(
\frac{X^{\prime}}{X}\frac{Y^{\prime}}{Y}\right)  ^{2}+\left(  \frac{X^{\prime
}}{X}\frac{Z^{\prime}}{Z}\right)  ^{2}+\left(  \frac{Y^{\prime}}{Y}%
\frac{Z^{\prime}}{Z}\right)  ^{2}\right]  ,
\end{align}
as it is observed $I_{3}$ is also defined as follows:%
\begin{equation}
I_{3}=I_{1}-2I_{2}+\frac{1}{3}R^{2},\label{I3}%
\end{equation}
this definition is only valid when $n=4.$

The gravitational entropy is defined as follows (see \cite{gron1}%
-\cite{gron2}):%
\begin{equation}
P^{2}=\frac{I_{3}}{I_{2}}=\frac{I_{1}-2I_{2}-\frac{1}{3}R^{2}}{I_{2}}%
=\frac{I_{1}}{I_{2}}+\frac{1}{3}\frac{R^{2}}{I_{2}}-2.\label{penrose}%
\end{equation}

\section{Lie Method.\label{LM}}

In this section we will study eq. (\ref{defeq}) i.e.
\begin{equation}
\dddot{\rho}=K_{1}\ddot{\rho}\frac{\dot{\rho}}{\rho}+K_{2}\frac{\dot{\rho}%
^{3}}{\rho^{2}}+K_{3}\dot{G}\rho^{2}+K_{4}G\dot{\rho}\rho,\label{lie}%
\end{equation}
where the $\left(  K_{i}\right)  _{i=1}^{4}$ are given by eq. (\ref{monica}),
through the Lie group method, i.e. in particular we seek the forms of $G(t)$
for which our field equations admit symmetries i.e. are integrable (see for
example \cite{Ibra}-\cite{TonyCas}).

Following the standard procedure we need to solve the next system of PDEs%
\begin{align}
\rho^{3}\xi_{\rho}  &  =0,\label{lie1}\\
\rho^{3}\xi_{\rho\rho}  &  =0,\label{lie2}\\
K_{1}\rho\eta-K_{1}\rho^{2}\eta_{\rho}-9\rho^{3}\xi_{t\rho}+3\rho^{3}%
\eta_{\rho\rho}  &  =0,\label{lie3}\\
-K_{1}\rho^{2}\eta_{t}+3\rho^{3}\eta_{t\rho}-3\rho^{3}\xi_{tt}  &
=0,\label{lie4}\\
K_{1}\rho^{2}\xi_{\rho\rho}+K_{2}\rho\xi_{\rho}-\rho^{3}\xi_{\rho\rho\rho}  &
=0,\label{lie5}\\
2K_{2}\rho\eta_{\rho}-K_{1}\rho^{2}\eta_{\rho\rho}+2K_{1}\rho^{2}\xi_{t\rho
}-2K_{2}\eta-3\rho^{3}\xi_{t\rho\rho}+\rho^{3}\eta_{\rho\rho\rho}  &
=0,\label{lie6}\\
3K_{2}\rho\eta_{t}-2K_{1}\rho^{2}\eta_{t\rho}+3K_{4}\rho^{4}G\xi_{\rho}%
+K_{1}\rho^{2}\xi_{tt}-3\rho^{3}\xi_{tt\rho}+3\rho^{3}\eta_{t\rho\rho}  &
=0,\label{lie7}\\
K_{4}\rho^{4}G^{\prime}\xi+K_{4}\rho^{3}G\eta+2K_{4}\rho^{4}G\xi_{t}-K_{1}%
\rho^{2}\eta_{tt}+3\rho^{3}\eta_{tt\rho}-\rho^{3}\xi_{tt\,t}-4K_{3}\rho
^{5}G^{\prime}\xi_{\rho}  &  =0,\label{lie8}\\
K_{4}\rho^{4}G\eta_{t}+\rho^{3}\eta_{tt\,t}+K_{3}\rho^{5}G^{\prime}\eta_{\rho
}-3K_{3}\rho^{5}G^{\prime}\xi_{t}-K_{3}\rho^{5}G^{\prime\prime}\xi-2K_{3}%
\rho^{4}G^{\prime}\eta &  =0,\label{lie9}%
\end{align}

To study all the possible solutions for this system of PDE may be very
tedious, for this reason we impose the symmetry $X=\left(  at+e\right)
\partial_{t}+b\rho\partial_{\rho},$ i.e. $\xi=at+e,\,\eta=b\rho,$ in such a
way that we get the following restrictions for $G(t)$. From eq. (\ref{lie8})
we get
\begin{equation}
\frac{G^{\prime}}{G}=-\frac{b+2a}{at+e},\label{rest1}%
\end{equation}
while from eq. (\ref{lie9}) it is obtained:%
\begin{equation}
\frac{G^{\prime\prime}}{G^{\prime}}=-\frac{3a+b}{at+e},\label{rest2}%
\end{equation}
where $a,b,e\in\mathbb{R}.$ Note that $\left[  a\right]  =\left[  b\right]
=1,$ i.e. they are dimensionless constants but $\left[  e\right]  =T$, with
respect to a dimensional base $B=\left\{  L,M,T\right\}  .$ So we are going to
study the different solutions that we may obtain in function of the constants
$a,b,e.$

\subsection{Scale symmetry.}

Making $e=0$ $i.e.$ considering only $\left(  \xi=at,~\eta=b\rho\right)  ,$ we
have to integrate eqs. (\ref{rest1} \ and \ref{rest2}), so
\begin{align}
\frac{G^{\prime}}{G} &  =-\frac{b+2a}{at},\qquad\Longrightarrow\qquad
G=G_{0}\left(  t\right)  ^{-\left(  2+\frac{b}{a}\right)  },\\
\frac{G^{\prime\prime}}{G^{\prime}} &  =-\frac{3a+b}{at},\qquad\Longrightarrow
\qquad G=C_{2}+G_{0}\left(  t\right)  ^{-\left(  2+\frac{b}{a}\right)  },
\end{align}
therefore we get
\begin{equation}
G=G_{0}\left(  t\right)  ^{-\left(  2+\frac{b}{a}\right)  },\label{G_scale}%
\end{equation}
where we assume that $G_{0}>0.$

The invariant solution for the energy density is:
\begin{equation}
\frac{bdt}{at}=\frac{d\rho}{\rho}\qquad\Longrightarrow\qquad\rho=\rho
_{0}t^{b/a},
\end{equation}
and for physical reasons we impose the condition, $ab<0$ then $b<0.$ We have
considered only the invariant solution since as we already know, the general
one usually lacks of any physical meaning as in this case (see for example
\cite{TonyCas}). Furthermore, as we will see in section \ref{SS}, this
spacetime is self-similar, this means that all the quantities must follow a
power law as in this case (see for example \cite{Wainwrit}-\cite{Jantzen}).

If we make that this solution verifies eq. (\ref{lie}) with $G(t)$ given by
eq. (\ref{G_scale}), we find the value of constant $\rho_{0},$ so%
\begin{equation}
\rho_{0}=-\frac{c^{2}b}{G_{0}12\pi a\left(  1+\omega\right)  ^{2}},
\end{equation}
with the only restriction $\omega\neq-1.$ Note that $ab<0,$ so $\rho_{0}>0.$

\begin{remark}
As we can see, it is verified the relationship $G\rho=t^{-2},$ i.e. the Mach
relationship for the inertia.
\end{remark}

Therefore, at this time we have the following behavior for $G(t)$
\begin{equation}
G(t)=G_{o}t^{-\left(  2+\frac{b}{a}\right)  },\qquad G\thickapprox\left\{
\begin{array}
[c]{l}%
\text{decreasing if }b/a\in\left(  -2,0\right)  ,\\
\text{constant if }b/a=-2,\\
\text{growing if }b/a<-2
\end{array}
\right.  .
\end{equation}
With these solutions we find the behavior of the rest of the quantities. From
eq.
\begin{equation}
\dot{\Lambda}=-\frac{8\pi}{c^{4}}\dot{G}\rho=\frac{8\pi}{c^{4}}\left(
2+\frac{b}{a}\right)  \rho_{0}G_{0}t^{-3},\Longrightarrow\Lambda=\Lambda
_{0}t^{-2},\qquad\Lambda_{0}=\frac{1}{3c^{2}\left(  1+\omega\right)  ^{2}%
}\frac{b}{a}\left(  2+\frac{b}{a}\right)  ,
\end{equation}
i.e.%
\begin{equation}
\Lambda(t)=\Lambda_{0}t^{-2},\qquad\Lambda_{0}\thickapprox\left\{
\begin{array}
[c]{l}%
\text{negative if }b/a\in\left(  -2,0\right)  ,\\
\text{vanish if }b/a=-2,\\
\text{positive if }b/a<-2
\end{array}
\right.  ,
\end{equation}
therefore, if $\Lambda_{0}>0,$ then $G$ is a growing time function but if
$\Lambda_{0}<0,$ then $G$ is decreasing.

Whit regard to $H$ we find that from eq. (\ref{eq14})%
\begin{equation}
H=-\left(  \frac{b}{a\left(  1+\omega\right)  }\right)  \frac{1}{t},\qquad
q=\frac{d}{dt}\left(  \frac{1}{H}\right)  -1=\frac{a\left(  1+\omega\right)
-b}{b},
\end{equation}
so
\begin{equation}
R=R_{0}\rho^{-1/3\left(  1+\omega\right)  }=R_{0}t^{-b/3a\left(
1+\omega\right)  },\qquad XYZ=R_{0}t^{-b/a\left(  1+\omega\right)  }.
\end{equation}

If we assume that the functions $\left(  X,Y,Z\right)  $ follow a power law
(i.e. $X=X_{0}t^{\alpha_{1}})$ then we get the following result%
\begin{equation}
Kt^{\alpha}=R_{0}t^{-b/a\left(  1+\omega\right)  },\qquad\Longrightarrow
\qquad\sum_{i}^{3}\alpha_{i}=\alpha=-\frac{b}{^{a\left(  1+\omega\right)  }}%
\end{equation}
where we may assume that $\left(  \alpha_{i}\right)  >0,\forall i$ and
$\left(  \alpha_{i}\neq\alpha_{j}\right)  $ although $\left(  \alpha
_{i}\rightarrow\alpha_{j}\right)  $ when $t\rightarrow\infty,$ with $i\neq j,$
but we have not more information about this behavior.

The shear is calculated as follows$.$%
\begin{equation}
\sigma^{2}=\frac{1}{3c^{2}\left(  1+\omega\right)  ^{2}}\left(  \frac
{\dot{\rho}}{\rho}\right)  ^{2}-\left(  8\pi\frac{G}{c^{4}}\rho+\Lambda
\right)  =\frac{1}{3c^{2}\left(  1+\omega\right)  ^{2}}\left(  \frac{b^{2}%
}{a^{2}}+2\frac{b}{a}-\frac{b}{a}\left(  2+\frac{b}{a}\right)  \right)
t^{-2}=0.
\end{equation}
therefore the shear vanish. In this way, this is an unphysical solution since
we are only interested in solutions which verify the condition $\sigma\neq0.$
Actually, we have obtained the flat FRW solution as we will show at the end of
this section.

Nevertheless if we follow the hypothesis of \ power law for the scale factors
$\left(  X,Y,Z\right)  $, then we get from the definition of the shear (see
eq. (\ref{defshear})) that it follows
\begin{equation}
\sigma^{2}=\frac{1}{6c^{2}}\left(  \sum_{i}^{3}\alpha_{i}^{2}-\sum_{i\neq
j}^{3}\alpha_{i}\alpha_{j}\right)  \frac{1}{t^{2}}\neq0,
\end{equation}
note that $\sigma\rightarrow0$ when $\left(  \alpha_{i}\rightarrow\alpha
_{j}\right)  .$ Note that at least one expect this behavior but as we will see
in the bellow subsection (full symmetry) actually $\sigma=0,$ since
$\alpha_{i}=\alpha_{j}$, (the flat FRW solution) so this is an unexpected result.

\subsection{Exponential behavior.}

Making $a=0,$ we have $\left(  \xi=e,\,\eta=b\rho\right)  ,$ so integrating
eqs. (\ref{rest1}-\ref{rest2}) it yields
\begin{align}
\frac{G^{\prime}}{G} &  =-\frac{b}{e},\Longrightarrow G=G_{0}\exp(-\frac{b}%
{e}t),\\
\frac{G^{\prime\prime}}{G^{\prime}} &  =-\frac{b}{e},\Longrightarrow
G=C_{2}+G_{0}\exp(-\frac{b}{e}t),
\end{align}
in this way we obtain
\begin{equation}
G=G_{0}\exp(-\frac{b}{e}t),
\end{equation}
where we assume that $G_{0}>0$

The invariant solution for the energy density is:
\begin{equation}
\frac{bdt}{e}=\frac{d\rho}{\rho}\qquad\Longrightarrow\qquad\rho=\rho_{0}%
\exp(\frac{b}{e}t)
\end{equation}
with the restriction, $eb<0$ with $b<0,$ from physical considerations. In
order to calculate the value of constant $\rho_{0},$ this solution must
verifies eq. (\ref{lie}) finding in this way that constant $\rho_{0},$ vanish
i.e. $\rho_{0}=0$. Note that with the obtained behavior for $G$ and $\rho,$
such functions must verify the field eq. (\ref{eq9}).

\subsection{Solution with the full symmetry.}

In this case we have the full symmetry i.e. $\left(  \xi=at+e,~\eta
=b\rho\right)  ,$ so, by integrating the constrains we find that%
\begin{equation}
\frac{G^{\prime}}{G}=-\frac{b+2a}{at+e}\Longrightarrow G=G_{0}%
(at+e)^{-(2a+b)/a},
\end{equation}%
\begin{equation}
\frac{G^{\prime\prime}}{G^{\prime}}=-\frac{3a+b}{at+e}\Longrightarrow
G=G_{0}(at+e)^{-(2a+b)/a},\label{G_KSS}%
\end{equation}
with $\left[  e\right]  =T,$ that is, we may interpret it as a characteristic
time, like in the ad hoc solution given by Kalligas et al (see \cite{We})$.$

The invariant solution for the energy density is%
\begin{equation}
\rho=\rho_{0}(at+e)^{b/a},
\end{equation}
where we need to impose the physical constrain such that $ab<0$ then $b<0,$ in
such a way that it must verifies eq. (\ref{lie}) with $G(t)$ given by eq.
(\ref{G_KSS}), finding the value of the numerical constant $\rho_{0},$ so%
\begin{equation}
\rho_{0}=-\frac{c^{2}b}{12\pi aG_{0}\left(  1+\omega\right)  ^{2}},
\end{equation}
we assume that $\omega\neq-1.$ As it is observed this is a nonsingular
solution since when $t\rightarrow0$ if $e\neq0,$ then $\rho\neq\infty.$

Therefore we have the following behavior for $G(t):$
\begin{equation}
G(t)=G_{0}(at+e)^{-(2a+b)/a},\qquad G\thickapprox\left\{
\begin{array}
[c]{l}%
\text{growing if }b/a\in\left(  -2,0\right)  ,\\
\text{constant if }b/a=-2,\\
\text{decreasing if }b/a<-2
\end{array}
\right.  .
\end{equation}

\begin{remark}
Note once again that it is verified the following relationship: $G\rho
\thickapprox(at+e)^{-2}.$
\end{remark}

We go next to calculate the rest of the quantities. From (\ref{eq9}) we get
\begin{equation}
\Lambda=\Lambda_{0}(at+e)^{-2},\qquad\Lambda_{0}=\left(  \frac{1}%
{6c^{2}\left(  1+\omega\right)  ^{2}}\frac{b\left(  2a+b\right)  }{a^{2}%
}\right)  ,
\end{equation}
i.e.%
\begin{equation}
\Lambda(t)=\Lambda_{0}(at+e)^{-2},\qquad\Lambda_{0}\thickapprox\left\{
\begin{array}
[c]{l}%
\text{negative if }\left\vert b\right\vert <2a,\\
\text{vanish if }\left\vert b\right\vert =2a,\\
\text{positive if }\left\vert b\right\vert >2a
\end{array}
\right.  .
\end{equation}

With regard to the quantity $H,$ we find form eq. (\ref{eq14}) that
\begin{equation}
R=R_{0}\rho^{-1/3\left(  1+\omega\right)  }=R_{0}(at+e)^{-b/3a\left(
1+\omega\right)  },
\end{equation}
and
\begin{equation}
H=-\left(  \frac{b}{a\left(  1+\omega\right)  }\right)  \frac{1}{t},\qquad
q=\frac{d}{dt}\left(  \frac{1}{H}\right)  -1=-\frac{a\left(  1+\omega\right)
}{b}-1,
\end{equation}
hence
\begin{equation}
XYZ=R_{0}(at+e)^{-b/a\left(  1+\omega\right)  },
\end{equation}
so (following the same argument as above) the functions $\left(  X,Y,Z\right)
$ follow a power law (i.e. $X=X_{0}(at+e)^{\alpha_{1}},$ etc...$)$ it is found
that%
\begin{equation}
K(at+e)^{\alpha}=R_{0}(at+e)^{-b/a\left(  1+\omega\right)  },\qquad
\Longrightarrow\qquad\sum_{i}^{3}\alpha_{i}=\alpha=-\frac{b}{^{a\left(
1+\omega\right)  }},\label{dorota}%
\end{equation}
where we may \textquotedblleft assume" that $\left(  \alpha_{i}\right)
>0,\forall i$ and $\left(  \alpha_{i}\neq\alpha_{j}\right)  $ although
$\left(  \alpha_{i}\rightarrow\alpha_{j}\right)  $ when $t\rightarrow\infty,$
and $i\neq j.$

The shear has the following behavior$.$%
\begin{equation}
\sigma^{2}=\frac{1}{3c^{2}\left(  1+\omega\right)  ^{2}}\left(  \frac
{\dot{\rho}}{\rho}\right)  ^{2}-\left(  8\pi\frac{G}{c^{4}}\rho+\Lambda
\right)  =\frac{1}{3c^{2}\left(  1+\omega\right)  ^{2}}\left(  b^{2}\left(
1-\frac{1}{2a^{2}}\right)  +\frac{b}{a}\right)  (at+e)^{-2}\neq0,
\end{equation}
finding that
\begin{equation}
b\left(  b\left(  2a^{2}-1\right)  +2a\right)  >0\Longleftrightarrow\qquad
b\left(  2a^{2}-1\right)  +2a<0,\Longleftrightarrow\left\vert b\left(
2a^{2}-1\right)  \right\vert >2a\qquad\Longrightarrow\left\{
\begin{array}
[c]{l}%
a>\sqrt{\frac{1}{2}}\\
\left\vert b\right\vert >\frac{2a}{2a^{2}-1}%
\end{array}
\right.  .
\end{equation}
since $\sigma_{0}^{2}$ must be positive.

At this point it seems that we have found a physical solution that depends on
the value of constants $a$ and $b.$ But, how to calculate the value of
constants $\left(  \alpha_{i}\right)  _{i=1}^{3}?.$

Now we will try to calculate the possible values for the constants $\left(
\alpha_{i}\right)  _{i=1}^{3},$ simply all these results must satisfy the
field eqs., so they are solution of the following system:
\begin{align}
\alpha_{1}\alpha_{2}+\alpha_{1}\alpha_{3}+\alpha_{2}\alpha_{3} &  =\frac{1}%
{6}\left(  \alpha^{2}+\frac{2\alpha}{\left(  1+\omega\right)  }\right)  ,\\
\alpha_{2}\left(  \alpha_{2}-1\right)  +\alpha_{3}\left(  \alpha_{3}-1\right)
+\alpha_{3}\alpha_{2} &  =\frac{1}{6}\left(  \alpha^{2}-\frac{2\alpha\left(
2\omega+1\right)  }{\left(  1+\omega\right)  }\right)  ,\\
\alpha_{1}\left(  \alpha_{1}-1\right)  +\alpha_{3}\left(  \alpha_{3}-1\right)
+\alpha_{3}\alpha_{1} &  =\frac{1}{6}\left(  \alpha^{2}-\frac{2\alpha\left(
2\omega+1\right)  }{\left(  1+\omega\right)  }\right)  ,\\
\alpha_{2}\left(  \alpha_{2}-1\right)  +\alpha_{1}\left(  \alpha_{1}-1\right)
+\alpha_{1}\alpha_{2} &  =\frac{1}{6}\left(  \alpha^{2}-\frac{2\alpha\left(
2\omega+1\right)  }{\left(  1+\omega\right)  }\right)  ,
\end{align}
and this system has only two solutions, the trivial one $\left(  \alpha
_{1}=\alpha_{2}=\alpha_{3}=0\right)  $ and the flat FRW one i.e.%
\begin{equation}
\alpha_{1}=\alpha_{2}=\alpha_{3}=\frac{2}{3\left(  1+\omega\right)  },
\end{equation}
which is incredible, this means, that from eq. (\ref{dorota}) we obtain $b=-2
$ and $a=1,$ but with \ these values for constants $\left(  a,b\right)  ,$ $G$
is a true \textquotedblleft constant\textquotedblright, $\Lambda$ vanishes and
$\sigma=0$. This solution was obtained by Einstein\&de Sitter (\cite{EdS}) in
1932 for $\omega=0$, and later by Harrison (\cite{Harrison}) $\forall\omega.$

We think that the followed method is too restrictive and for this reason we
obtain only this solution.

Therefore this method brings us to obtain only the classical flat FRW solution
with $G$ constant and $\Lambda$ equal nought.

\section{Self-similar solution.\label{SS}}

In general relativity, the term self-similarity can be used in two ways. One
is for the properties of spacetimes, the other is for the properties of matter
fields. These are not equivalent in general. The self-similarity in general
relativity was defined for the first time by Cahill and Taub (see \cite{CT},
and for general reviews \cite{21}-\cite{Hall}). Self-similarity is defined by
the existence of a homothetic vector ${V}$ in the spacetime, which satisfies
\begin{equation}
\mathcal{L}_{V}g_{ij}=2\alpha g_{ij},\label{gss1}%
\end{equation}
where $g_{ij}$ is the metric tensor, $\mathcal{L}_{V}$ denotes Lie
differentiation along ${V}$ and $\alpha$ is a constant. This is a special type
of conformal Killing vectors. This self-similarity is called homothety. If
$\alpha\neq0$, then it can be set to be unity by a constant rescaling of ${V}%
$. If $\alpha=0$, i.e. $\mathcal{L}_{V}g_{ij}=0$, then ${V}$ is a Killing vector.

Homothety is a purely geometric property of spacetime so that the physical
quantity does not necessarily exhibit self-similarity such as $\mathcal{L}%
_{V}Z=dZ$, where $d$ is a constant and $Z$ is, for example, the pressure, the
energy density and so on. From equation (\ref{gss1}) it follows that
\begin{equation}
\mathcal{L}_{V}R^{i}\,_{jkl}=0,
\end{equation}
and hence
\begin{equation}
\mathcal{L}_{V}R_{ij}=0,\qquad\mathcal{L}_{V}G_{ij}=0.\label{mattercoll}%
\end{equation}
A vector field ${V}$ that satisfies the above equations is called a curvature
collineation, a Ricci collineation and a matter collineation, respectively. It
is noted that such equations do not necessarily mean that ${V}$ is a
homothetic vector. We consider the Einstein equations
\begin{equation}
G_{ij}=8\pi GT_{ij},\label{einstein}%
\end{equation}
where $T_{ij}$ is the energy-momentum tensor.

If the spacetime is homothetic, the energy-momentum tensor of the matter
fields must satisfy
\begin{equation}
\mathcal{L}_{V}T_{ij}=0,\label{emcoll}%
\end{equation}
through equations~(\ref{einstein}) and (\ref{mattercoll}). For a perfect fluid
case, the energy-momentum tensor takes the form of eq. (\ref{eq0}) i.e.
$T_{ij}=(p+\rho)u_{i}u_{j}+pg_{ij},$where $p$ and $\rho$ are the pressure and
the energy density, respectively. Then, equations~(\ref{gss1}) and
(\ref{emcoll}) result in
\begin{equation}
\mathcal{L}_{V}u^{i}=-\alpha u^{i},\qquad\mathcal{L}_{V}\rho=-2\alpha
\rho,\qquad\mathcal{L}_{V}p=-2\alpha p.\label{ssmu}%
\end{equation}
As shown above, for a perfect fluid, the self-similarity of the spacetime and
that of the physical quantity coincide. However, this fact does not
necessarily hold for more general matter fields. Thus the self-similar
variables can be determined from dimensional considerations in the case of
homothety. Therefore, we can conclude homothety as the general relativistic
analogue of complete similarity.

From the constraints (\ref{ssmu}), we can show that if we consider the
barotropic equation of state, i.e., $p=f(\rho)$, then the equation of state
must have the form $p=\omega\rho$, where $\omega$ is a constant. This class of
equations of state contains a stiff fluid ($\omega=1$) as special cases,
whiting this theoretical framework. There are many papers devoted to study
Bianchi I models (in different context) assuming the hypothesis of
self-similarity (see for example \cite{HW}-\cite{griego}) but here, we would
like to try to show how taking into account this class of hypothesis one is
able to find exact solutions to the field equations within the framework of
the time varying constants.

The homothetic equations are given by eq. (\ref{gss1}) so it is a
straightforward task to find the homothetic vector field, where in this case
is as follows:%
\begin{equation}
V=t\partial_{t}+\left(  1-t\frac{\dot{X}}{X}\right)  x\partial_{x}+\left(
1-t\frac{\dot{Y}}{Y}\right)  y\partial_{y}+\left(  1-t\frac{\dot{Z}}%
{Z}\right)  z\partial_{z},\label{HO1}%
\end{equation}
iff the following ODE is satisfied
\begin{equation}
\left(  X\dot{X}+tX\ddot{X}-t\left(  \dot{X}\right)  ^{2}\right)
x=0,\label{helen}%
\end{equation}
and so on with respect to $\left(  Y,y\right)  $ and $\left(  Z,z\right)  .$

As it is observed from eq. (\ref{helen}) if we simplify this ODE it is
obtained the following one:%
\begin{equation}
\frac{H_{1}^{\prime}}{H_{1}}=-\frac{1}{t}\Longleftrightarrow tH_{1}%
=const.\Longleftrightarrow X=X_{0}t^{\alpha_{1}},
\end{equation}
with $X_{0}$ and $\alpha\in\mathbb{R},$ etc....with respect to the others
scale factors. Note that $^{\prime}:=\frac{d}{dt}:=dot.$ i.e. $X^{\prime}%
=\dot{X}.$

Therefore, we have obtained the following behavior for the scale factors:%
\begin{equation}
X=X_{0}t^{\alpha_{1}},\qquad Y=Y_{0}t^{\alpha_{2}},\qquad Z=Z_{0}t^{\alpha
_{3}},
\end{equation}
with $X_{0},Y_{0},Z_{0}$ are integrating constants and $\left(  \alpha
_{i}\right)  _{i=1}^{3}\in\mathbb{R}.$ In this way we find that%
\begin{equation}
H=\left(  \sum_{i=1}^{3}\alpha_{i}\right)  \frac{1}{t}=\left(  \alpha
_{1}+\alpha_{2}+\alpha_{3}\right)  \frac{1}{t}=\frac{\alpha}{t},
\end{equation}
and hence%
\begin{equation}
\rho=\rho_{0}t^{-(\omega+1)\alpha}.
\end{equation}

From the field equations (\ref{eq3}) and (\ref{eq9}) we get that%
\begin{equation}
\Lambda^{\prime}=-\frac{A}{c^{2}}\frac{2}{t^{3}}-\frac{8\pi\rho_{0}}{c^{4}%
}\left(  G^{\prime}t^{-(\omega+1)\alpha}-(\omega+1)\alpha Gt^{-(\omega
+1)\alpha-1}\right)  ,\label{laura1}%
\end{equation}
where $A=\sum_{i\neq j}\alpha_{i}\alpha_{j}=\alpha_{1}\alpha_{2}+\alpha
_{3}\alpha_{1}+\alpha_{2}\alpha_{3},\alpha=\sum_{i}\alpha_{i}=\alpha
_{1}+\alpha_{2}+\alpha_{3},$ therefore
\begin{equation}
G=\frac{c^{2}}{4\pi\rho_{0}}\frac{A}{(\omega+1)\alpha}t^{(\omega+1)\alpha-2},
\end{equation}
as we can see it is verified the relationship $G\rho\thickapprox t^{-2},$ as
it is expected.

The behavior of $G$ is the following one:%
\begin{equation}
G=G_{0}t^{(\omega+1)\alpha-2}\qquad\Longrightarrow\qquad\left\{
\begin{array}
[c]{l}%
\text{growing if }(\omega+1)\alpha>2\\
\text{constant if }(\omega+1)\alpha=2\\
\text{decreasing if }(\omega+1)\alpha<2
\end{array}
\right.  ,
\end{equation}
with $G_{0}>0.$

Now, we go next to calculate the quantity $\Lambda,$ from eq. (\ref{laura1})
we get%
\begin{equation}
\Lambda=\Lambda_{0}t^{-2},\qquad\Lambda_{0}=\frac{A}{c^{2}}\left(  1-\frac
{2}{(\omega+1)\alpha}\right)  ,
\end{equation}
in this way it is observed that%
\begin{equation}
\Lambda=\Lambda_{0}t^{-2},\qquad\left\{
\begin{array}
[c]{c}%
\Lambda_{0}>0\Longleftrightarrow(\omega+1)\alpha>2\\
\Lambda_{0}=0\Longleftrightarrow(\omega+1)\alpha=2\\
\Lambda_{0}<0\Longleftrightarrow(\omega+1)\alpha<2
\end{array}
\right.  .
\end{equation}

The shear behaves (see eq. (\ref{defshear})) as follows:
\begin{equation}
\sigma^{2}=\frac{1}{3c^{2}}\left(  \sum_{i}^{3}\alpha_{i}^{2}-\sum_{i\neq
j}^{3}\alpha_{i}\alpha_{j}\right)  \frac{1}{t^{2}}\neq0,
\end{equation}
by hypothesis, since at this point we have not any information about the value
of the constants $\left(  \alpha_{i}\right)  _{i=1}^{3}.$

In order to find the value of constants $\left(  \alpha_{i}\right)  ,$ we make
that them verify the field eqs. so in this case we get the following system of
eqs.:%
\begin{align}
\alpha_{2}\left(  \alpha_{2}-1\right)  +\alpha_{3}\left(  \alpha_{3}-1\right)
+\alpha_{3}\alpha_{2} &  =A\left(  \frac{\alpha-2}{\alpha}\right)  ,\\
\alpha_{1}\left(  \alpha_{1}-1\right)  +\alpha_{3}\left(  \alpha_{3}-1\right)
+\alpha_{3}\alpha_{1} &  =A\left(  \frac{\alpha-2}{\alpha}\right)  ,\\
\alpha_{2}\left(  \alpha_{2}-1\right)  +\alpha_{1}\left(  \alpha_{1}-1\right)
+\alpha_{1}\alpha_{2} &  =A\left(  \frac{\alpha-2}{\alpha}\right)  ,
\end{align}
where $A=\alpha_{1}\alpha_{2}+\alpha_{3}\alpha_{1}+\alpha_{2}\alpha_{3},$ and
$\alpha=\alpha_{1}+\alpha_{2}+\alpha_{3}.$

So we have the following solutions for this system of equations:%
\begin{align}
\alpha_{1} &  =\alpha_{2}=\alpha_{3},\label{solsys1}\\
\alpha_{1} &  =1-\alpha_{2}-\alpha_{3},\label{solsys3}%
\end{align}
as it is observed solution (\ref{solsys1}) is not interesting for us, since it
is again the flat FRW solution. Only the second solution has physical meaning
(in this framework). Nevertheless we have found that this solution only
verifies the first of the condition of the Kasner like solutions i.e.
\begin{equation}
\alpha=\sum\alpha_{i}=1,
\end{equation}
while the second condition
\begin{equation}
\sum\alpha_{i}^{2}=1,
\end{equation}
it is not verified (see \cite{Kasner} \ and \ \cite{SH}). In this case we find
that it is verified the condition
\begin{equation}
\sum\alpha_{i}^{2}<1.
\end{equation}

Therefore we have found a similar behavior as the obtained one in (\cite{HW},
we say similar because there the authors only study standard models i.e.
models where the \textquotedblleft constants\textquotedblright\ are true
constants, in fact $\Lambda=0$), except than here this result is valid for all
equation of state i.e. $\forall\omega\in\left(  -1,1\right)  .$ Nevertheless
in reference (\cite{griego}), the authors claim that the solution must verify
both conditions, i.e. $\sum\alpha_{i}=1=\sum\alpha_{i}^{2}.$ See the end of
the kinematical self-similar solution for a comment on this class of solutions
as well as the appendix (\ref{ApB}).

Therefore we have obtained the following behavior for the main quantities:%
\begin{equation}
H=\frac{1}{t},\qquad\Longrightarrow\qquad q=0,
\end{equation}
so it is quite difficult to reconcile this model with the observational data.
With regard to the energy density we find that%
\begin{equation}
\rho=\rho_{0}t^{-\left(  \omega+1\right)  },\text{ }%
\end{equation}
\ so, if $\omega<-1\Longrightarrow\rho$ is growing (phantom cosmologies), for
the rest of the values of $\omega,$ i.e. $\omega\in(-1,1],$ $\rho$ is a
decreasing function on time.

$G$ has now a more restrictive behavior since (we are supposing that
$\omega\in(-1,1],$ i.e. $\omega>1,$ is forbidden)
\begin{equation}
G=G_{0}t^{\left(  \omega+1\right)  -2},\qquad\Longrightarrow\qquad\left\{
\begin{array}
[c]{l}%
\text{constant if }(\omega=1)\\
\text{decreasing }\forall\text{ }\omega\in(-1,1)
\end{array}
\right.  ,
\end{equation}
therefore we find that $G$ is a decreasing function on time. The cosmological
\textquotedblleft constant\textquotedblright\ behaves as follows
\begin{equation}
\Lambda=\Lambda_{0}t^{-2},\qquad\Lambda_{0}=\left\{
\begin{array}
[c]{l}%
\Lambda_{0}=0\Longleftrightarrow\omega=1\\
\Lambda_{0}<0,\qquad\forall\text{ }\omega\in(-1,1)
\end{array}
\right.  ,
\end{equation}
so we have found that $\Lambda$ is a negative decreasing function on time.

As we can see, this solution is quite similar to the obtained one through the
Lie method with the scaling symmetry, at least in order of magnitude.

With regard to the curvature behavior, we may see that
\begin{equation}
I_{1}=\frac{K}{c^{4}t^{4}},\qquad I_{2}=\frac{4(-\alpha_{2}-\alpha_{3}%
+\alpha_{2}^{2}+\alpha_{2}\alpha_{3}+\alpha_{3}^{2})^{2}}{c^{4}t^{4}},
\end{equation}
where $K=K(\alpha_{i})=const\neq0,$ i.e.%
\begin{equation}
K=\left[  3\left(  \alpha_{2}^{2}+\alpha_{3}^{2}\right)  +2\alpha_{2}%
\alpha_{3}+9\alpha_{2}^{2}\alpha_{3}^{2}+3\left(  \alpha_{2}^{4}+\alpha
_{3}^{4}\right)  -6\left(  \alpha_{2}^{3}+\alpha_{3}^{3}\right)  +\alpha
_{2}\alpha_{3}\left(  6\left(  \alpha_{2}^{2}+\alpha_{3}^{2}\right)  -8\left(
\alpha_{2}+\alpha_{3}\right)  \right)  \right]  .
\end{equation}

The non-zero components of the Weyl tensor are:%
\begin{align}
C_{1212} &  =K_{1}t^{-2(\alpha_{2}+\alpha_{3})},\qquad C_{1313}=K_{2}%
t^{-2(1-\alpha_{2})},\qquad C_{1414}=K_{3}t^{-2(1-\alpha_{3})},\nonumber\\
C_{2323} &  =K_{4}t^{-2\alpha_{3}},\qquad C_{2424}=K_{5}t^{-2\alpha_{2}%
},\qquad C_{3434}=K_{6}t^{-2(1-\alpha_{3}-\alpha_{2})},
\end{align}
where $\left(  K_{i}\right)  _{i=1}^{6}=K(\alpha_{i})=const\neq0,$ and taking
into account a very famous result by Hall et al (see \cite{hrv}, and the next
section \ref{MC}) we may check that
\begin{equation}
\mathcal{L}_{V}C_{jkl}^{i}=0,
\end{equation}
as it is shown in (\cite{hrv}) if a vector field $V\in\mathfrak{X}(M),$
verifies the conditions $\mathcal{L}_{V}C_{jkl}^{i}=0,$ and $\mathcal{L}%
_{V}T_{ij}=0,$ then $\mathcal{L}_{V}g=2g$ i.e. it is a homothetic vector
field, but in this case we have arrived to the conclusion that $\mathcal{L}%
_{V}g=2g\Longleftrightarrow\mathcal{L}_{V}T_{ij}=0$ (as we will see in the
next section) and that it is also verified the relationship $\mathcal{L}%
_{V}C_{jkl}^{i}=0.$

The non-zero components of the electric part of the Weyl tensor are:%
\begin{equation}
E_{22}=\tilde{K}_{1}t^{-2(\alpha_{2}+\alpha_{3})},\qquad E_{33}=\tilde{K}%
_{2}t^{-2(1-\alpha_{2})},\qquad E_{44}=\tilde{K}_{3}t^{-2(1-\alpha_{3})},
\end{equation}
and the last invariant has the following behavior%
\begin{equation}
I_{3}=\frac{\hat{K}}{c^{4}t^{4}},
\end{equation}
with $\hat{K}$ given by%
\begin{equation}
\hat{K}=\frac{16}{3}\left[  \alpha_{2}^{2}+\alpha_{3}^{2}-\alpha_{2}\alpha
_{3}+3\alpha_{2}^{2}\alpha_{3}^{2}+\alpha_{2}^{4}+\alpha_{3}^{4}-2\left(
\alpha_{2}^{3}+\alpha_{3}^{3}\right)  +\alpha_{2}\alpha_{3}\left(  2\left(
\alpha_{2}^{2}+\alpha_{3}^{2}\right)  -\left(  \alpha_{2}+\alpha_{3}\right)
\right)  \right]  ,
\end{equation}
while the gravitational entropy behaves as%
\begin{equation}
P^{2}=const.\neq0,
\end{equation}
note that $P^{2}=I_{3}/I_{2}.$

\section{Matter collineations.\label{MC}}

In recent years, much interest has been shown in the study of matter
collineation (MCs) (see for example \cite{Sharif}-\cite{TA}. A vector field
along which the Lie derivative of the energy-momentum tensor vanishes is
called an MC, i.e.%
\begin{equation}
{\mathcal{L}}_{V}T_{ij}=0,
\end{equation}
where $V^{i}$ is the symmetry or collineation vector. The MC equations, in
component form, can be written as
\begin{equation}
T_{ij,k}V^{k}+T_{ik}V_{,j}^{k}+T_{kj}V_{,i}^{k}=0,
\end{equation}
where the indices $i,j,k$ run from $0$ to $3$. Also, assuming the Einstein
field equations, a vector $V^{i}$ generates an MC if ${\mathcal{L}}_{V}%
G_{ij}=0$. It is obvious that the symmetries of the metric tensor (isometries)
are also symmetries of the Einstein tensor $G_{ij}$, but this is not
necessarily the case for the symmetries of the Ricci tensor (Ricci
collineations) which are not, in general, symmetries of the Einstein tensor.
If $V$ is a Killing vector (KV) (or a homothetic vector), then ${\mathcal{L}%
}_{V}T_{ij}=0$, thus every isometry is also an MC but the converse is not
true, in general. Notice that collineations can be proper (non-trivial) or
improper (trivial). Proper MC is defined to be an MC which is not a KV, or a
homothetic vector.

Carot et al (see \cite{ccv}) and Hall et al.(see \cite{hrv}) have noticed some
important general results about the Lie algebra of MCs.

Let $M$ be a spacetime manifold. Then, generically, any vector field $V$ on
$M$ which simultaneously satisfies ${\mathcal{L}}_{V}T_{ab}=0$
($\Leftrightarrow{\mathcal{L}}_{V}G_{ab}=0$) and ${\mathcal{L}}_{V}C_{bcd}%
^{a}=0$ is a homothetic vector field.

If $V$ is a Killing vector (KV) (or a homothetic vector), then ${\mathcal{L}%
}_{V}T_{ab}=0$, thus every isometry is also an MC but the converse is not
true, in general. Notice that collineations can be proper (non-trivial) or
improper (trivial). Proper MC is defined to be an MC which is not a KV, or a
homothetic vector.

Since the ST is SS then we already know that the SS vector field is also
matter collineation i.e. we would like to explore how such symmetries allow us
to obtain relationships between the quantities in such a way that it is not
necessary to make any hypothesis to a solution to the field equations. In
order to do that we need to modify the usual MC equations since with the usual
one we are not able to obtain information about the behavior of $G$ and
$\Lambda.$

\subsection{The usual matter collineation equations.}

For a vector field $V=\left(  V_{i}\left(  t,x,y,z\right)  \right)  _{i=1}%
^{4},$ the matter collineations reads:%
\begin{equation}
{\mathcal{L}}_{V}T_{ij}=0,
\end{equation}
so outlining the equations and integrating them (it is a straightforward task)
we obtain the following interesting relationships
\begin{equation}
\frac{\rho^{\prime}}{\rho}=-2\frac{V_{1}^{\prime}}{V_{1}},\qquad
\Longrightarrow\rho=K_{0}V_{1}^{-2}.
\end{equation}
and
\begin{equation}
V_{1}\left(  \frac{p^{\prime}}{p}+2H_{1}\right)  =-2\partial_{x}V_{2}%
,\qquad\Longrightarrow\qquad V_{1}\left(  \frac{\rho^{\prime}}{\rho}%
+2H_{1}\right)  =-2\partial_{x}V_{2},\qquad\Longrightarrow\qquad V_{2}%
=-V_{1}\left(  H_{1}-\frac{V_{1}^{\prime}}{V_{1}}\right)  x,
\end{equation}
in this way we find that%
\begin{equation}
V_{3}=-V_{1}\left(  H_{1}-\frac{V_{1}^{\prime}}{V_{1}}\right)  y,\qquad
V_{4}=-V_{1}\left(  H_{1}-\frac{V_{1}^{\prime}}{V_{1}}\right)  z,
\end{equation}
therefore the obtained matter collineation (MC) vector field is:%
\begin{equation}
V=V_{1}\partial_{t}+V_{1}\left(  \frac{V_{1}^{\prime}}{V_{1}}-H_{1}\right)
x\partial_{x}+V_{1}\left(  \frac{V_{1}^{\prime}}{V_{1}}-H_{2}\right)
y\partial_{y}+V_{1}\left(  \frac{V_{1}^{\prime}}{V_{1}}-H_{3}\right)
z\partial_{z},\label{mc}%
\end{equation}
where as it is observed if $V_{1}=t$, then it is obtained the homothetic
vector field (see eq. (\ref{HO1})) as we already know.

For this reason we may also check that the homothetic vector field verify the
relationship as well
\begin{equation}
L_{HO}T_{ij}=0,
\end{equation}
iff the following ODEs are satisfied:%
\begin{align}
c^{2}\left(  \rho^{\prime}t+2\rho\right)   &  =0,\qquad X^{2}\left(
p^{\prime}t+2p\right)  =0,\label{mc1}\\
-px\left(  XX^{\prime}+tXX^{\prime\prime}-t\left(  X^{\prime}\right)
^{2}\right)   &  =0,\label{mc2}%
\end{align}
as we can see the ODEs (\ref{mc1}) $\left(  \rho^{\prime}t+2\rho\right)  =0,$
and $p^{\prime}t+2p=0$ brings us to obtain $p=\omega\rho\thickapprox t^{-2},$
as it is expected in a model without time varying constant and from the eq.
$XX^{\prime}+tX^{\prime\prime}-t\left(  X^{\prime}\right)  ^{2}$ i.e.
$H_{1}+tH_{1}^{\prime}=0,$ that $tH_{1}=const.$ and so on with respect to the
scale factors $Y$ and $Z$ i.e. $tH_{i}=const,$ $i=2,3.$

Since we have not any information about the behavior of $\ G$ and $\Lambda,$
we \textquotedblleft suggest\textquotedblright\ the following modification.

\subsection{Modified matter collineations equations}

The problem arises when we are considering a model with $G$ and $\Lambda$
variable, so we suggest to change the above procedure i.e. the standard one
by
\begin{equation}
L_{V}\left(  \frac{G(t)}{c^{4}}T_{ij}\right)  =0.
\end{equation}

As in the above case it is easily outlined the resulting matter collineations
eqs. so it is a straightforward task to integrate them. In this case we have
found the following relationships between the quantities:
\begin{equation}
\frac{G^{\prime}}{G}+\frac{\rho^{\prime}}{\rho}=-2\frac{V_{1}^{\prime}}{V_{1}%
},\qquad\Longrightarrow G\rho=K_{0}V_{1}^{-2}.
\end{equation}
where $K_{0}$ is an integration constant, and as above we find that%
\begin{equation}
V_{2}=-V_{1}\left(  H_{1}-\frac{V_{1}^{\prime}}{V_{1}}\right)  x,\qquad
V_{3}=-V_{1}\left(  H_{1}-\frac{V_{1}^{\prime}}{V_{1}}\right)  y,\qquad
V_{4}=-V_{1}\left(  H_{1}-\frac{V_{1}^{\prime}}{V_{1}}\right)  z,
\end{equation}
obtaining again
\begin{equation}
V=V_{1}\partial_{t}+V_{1}\left(  \frac{V_{1}^{\prime}}{V_{1}}-H_{1}\right)
x\partial_{x}+V_{1}\left(  \frac{V_{1}^{\prime}}{V_{1}}-H_{2}\right)
y\partial_{y}+V_{1}\left(  \frac{V_{1}^{\prime}}{V_{1}}-H_{3}\right)
z\partial_{z},\label{nmc}%
\end{equation}
but in this case, it is verified the relationship
\begin{equation}
G\rho=K_{0}V_{1}^{-2}.
\end{equation}

As in the previous case we may check that its is verified the equation
$L_{HO}\left(  \frac{G(t)}{c^{4}}T_{ij}\right)  =0.$%
\begin{align}
\frac{8\pi}{c^{2}}\left(  t\rho G^{\prime}+tG\rho^{\prime}+2G\rho\right)   &
=0,\qquad\Longleftrightarrow\qquad\frac{G^{\prime}}{G}+\frac{\rho^{\prime}%
}{\rho}=-\frac{2}{t}\Longleftrightarrow G\rho\thickapprox t^{-2},\\
-px\left(  XX^{\prime}+tXX^{\prime\prime}-t\left(  X^{\prime}\right)
^{2}\right)   &  =0,\qquad\Longleftrightarrow\qquad tH_{1}^{\prime}%
=-H_{1}\qquad\Longleftrightarrow X=X_{0}t^{\alpha_{1}},\\
\frac{8\pi}{c^{4}}X^{2}\left(  tpG^{\prime}+tGp^{\prime}+2pG\right)   &
=0,\qquad\Longleftrightarrow\qquad\frac{G^{\prime}}{G}+\frac{p^{\prime}}%
{p}=-\frac{2}{t}\Longleftrightarrow Gp\thickapprox t^{-2},
\end{align}
from this equations we also arrive to the obvious conclusions that
$p=\omega\rho$ with $\omega\in\mathbb{R}.$

But unfortunately we have not any information about the behavior of the
cosmological constant $\Lambda$ for this reason we suggest the following modification.

\subsection{The complete modification of MC equations.}

In this case we consider that must be satisfied the following equation%
\begin{equation}
L_{V}\left(  \frac{G(t)}{c^{4}}T_{ij}-\Lambda(t)g_{ij}\right)  =0,
\end{equation}
that we reinterprets as%
\begin{equation}
L_{V}\left(  \frac{G(t)}{c^{4}}T_{ij}\right)  =0=L_{V}\left(  \Lambda
(t)g_{ij}\right)  .
\end{equation}

As in the above cases (we are following the same procedure in both cases) we
find again that again
\begin{equation}
V=V_{1}\partial_{t}+V_{1}\left(  \frac{V_{1}^{\prime}}{V_{1}}-H_{1}\right)
x\partial_{x}+V_{1}\left(  \frac{V_{1}^{\prime}}{V_{1}}-H_{2}\right)
y\partial_{y}+V_{1}\left(  \frac{V_{1}^{\prime}}{V_{1}}-H_{3}\right)
z\partial_{z},
\end{equation}

Therefore checking the relationship
\begin{equation}
L_{HO}\left(  \frac{G(t)}{c^{4}}T_{ij}-\Lambda(t)g_{ij}\right)  =0.
\end{equation}
it is obtained the following results
\begin{align}
\frac{8\pi}{c^{4}}\left(  t\rho G^{\prime}+tG\rho^{\prime}+2G\rho\right)   &
=-\left(  t\Lambda^{\prime}+2\Lambda\right)  ,\qquad\Longleftrightarrow
\qquad\frac{G^{\prime}}{G}+\frac{\rho^{\prime}}{\rho}=-\frac{2}{t}%
\Longleftrightarrow G\rho\thickapprox t^{-2},\\
\left(  -\frac{8\pi G}{c^{4}}p+\Lambda\right)  x\left(  XX^{\prime
}+tXX^{\prime\prime}-t\left(  X^{\prime}\right)  ^{2}\right)   &
=0,\qquad\qquad\qquad\Longleftrightarrow\qquad tH_{1}^{\prime}=-H_{1}%
\qquad\Longleftrightarrow X=X_{0}t^{\alpha_{1}},\\
\left(  -\frac{8\pi G}{c^{4}}p+\Lambda\right)  y\left(  YY^{\prime
}+tYY^{\prime\prime}-t\left(  Y^{\prime}\right)  ^{2}\right)   &
=0,\qquad\qquad\qquad\Longleftrightarrow\qquad tH_{2}^{\prime}=-H_{2}%
\qquad\Longleftrightarrow Y=Y_{0}t^{\alpha_{2}},\\
\left(  -\frac{8\pi G}{c^{4}}p+\Lambda\right)  z\left(  ZZ^{\prime
}+tZZ^{\prime\prime}-t\left(  Z^{\prime}\right)  ^{2}\right)   &
=0,\qquad\qquad\qquad\Longleftrightarrow\qquad tH_{3}^{\prime}=-H_{3}%
\qquad\Longleftrightarrow Z=Z_{0}t^{\alpha_{3}},\\
\frac{8\pi}{c^{4}}\left(  tpG^{\prime}+tGp^{\prime}+2pG\right)   &  =\left(
t\Lambda^{\prime}+2\Lambda\right)  ,\qquad\Longleftrightarrow\qquad
\frac{G^{\prime}}{G}+\frac{p^{\prime}}{p}=-\frac{2}{t}\Longleftrightarrow
Gp\thickapprox t^{-2},
\end{align}
while
\begin{equation}
\Lambda=\Lambda_{0}t^{-2},
\end{equation}
that is to say, we have obtained the same results than in the SS section (as
it was expected).

\begin{remark}
Other possibilities could be explored, for example to calculate the algebra of
the matter tensor field $T\in T_{1}^{1}(M)$ as well as $T\in T_{0}^{2}(M), $
note that here we only have studied the case $T\in T_{2}^{0}(M).$ Nevertheless
and unfortunately for these cases we have not been able to find any
interesting result i.e. with physical meaning. I would to thank Prof. G. Hall
for drawing my attention about this fact.
\end{remark}

\section{Kinematic Self-similarity.\label{KSS}}

Kinematic self-similarity has been defined in the context of relativistic
fluid mechanics as an example of incomplete similarity (see for example
\cite{CH1}-\cite{Sintes-KSS}). It should be noted that the introduction of
incomplete similarity to general relativity is not unique.

A spacetime is said to be kinematic self-similar if it admits a kinematic
self-similar vector ${V}$ which satisfies the conditions
\begin{align}
{\mathcal{L}}_{V}h_{ij} &  =2\delta h_{ij},\label{kss}\\
{\mathcal{L}}_{V}u_{i} &  =\alpha u_{i},\label{gss}%
\end{align}
where $u^{i}$ is the four-velocity of the fluid and $h_{ij}=g_{ij}+u_{i}u_{j}
$ is the projection tensor, and $\alpha$ and $\delta$ are constants~.

If $\delta\neq0$, the similarity transformation is characterized by the
scale-independent ratio $\alpha/\delta$, which is referred to as the
similarity index. If the ratio is unity, ${V}$ turns out to be a homothetic
vector. In the context of kinematic self-similarity, homothety is referred to
as self-similarity of the first kind. If $\alpha=0$ and $\delta\neq0$, it is
referred to as self-similarity of the zeroth kind. If the ratio is not equal
to zero or one, it is referred to as self-similarity of the second kind. If
$\alpha\neq0$ and $\delta=0$, it is referred to as self-similarity of the
infinite kind. If $\delta=\alpha=0$, ${V}$ turns out to be a Killing vector.

From the Einstein equation (\ref{einstein}), we can derive
\begin{equation}
\mathcal{L}_{V}G_{ij}=8\pi G\mathcal{L}_{V}T_{ij},\label{intcondition}%
\end{equation}
this equation is called the integrability condition.

When a perfect fluid is irrotational, i.e., $\omega_{ij}=0$, the Einstein
equations and the integrability conditions (\ref{intcondition}) give%
\begin{equation}
(\alpha-\delta)\mathcal{R}_{ij}=0,
\end{equation}
where $\mathcal{R}_{ij}$ is the Ricci tensor on the hypersurface orthogonal to
$u^{i}$. This means that if a solution is kinematic self-similar but not
homothetic and if the fluid is irrotational, then the hypersurface orthogonal
to fluid flow is flat.

From the physical point of view the detailed study of cosmological models
admitting KSS shows that they can represent asymptotic states of more general
models or, under certain conditions, they are asymptotic to an exact
homothetic solution \cite{Coley-KSS,Benoit-Coley}.

Therefore and following the same idea as in the above sections we would like
to extend this hypothesis in order to find exact solutions to cosmological
models with time varying constant.

Kinematic self-similarity are characterized by the equations (\ref{kss}%
-\ref{gss}), so in this way it is found that the vector field $V:=KSS$ is:%
\begin{equation}
KSS=-(\alpha t+\beta)\partial_{t}+f_{1}x\partial_{x}+f_{2}y\partial_{y}%
+f_{3}z\partial_{z},
\end{equation}
where%
\begin{equation}
f_{1}=\left(  \delta+(\alpha t+\beta)\frac{\dot{X}}{X}\right)  ,\quad
f_{2}=\left(  \delta+(\alpha t+\beta)\frac{\dot{Y}}{Y}\right)  ,\quad
f_{3}=\left(  \delta+(\alpha t+\beta)\frac{\dot{Z}}{Z}\right)  .
\end{equation}

As in the case of the homothetic vector field in this case it is necessary to
satisfy the following ODE
\begin{equation}
-\alpha\frac{X^{\prime}}{X}-\left(  \alpha t+\beta\right)  \frac
{X^{\prime\prime}}{X}+\left(  \alpha t+\beta\right)  \left(  \frac{X^{\prime}%
}{X}\right)  ^{2}=0,
\end{equation}
arriving to the same conclusion i.e.%
\begin{equation}
H_{1}=\frac{1}{\left(  \alpha t+\beta\right)  },
\end{equation}
and therefore the solution follows a power law and hence%
\begin{equation}
X=X_{0}\left(  t+\frac{\beta}{\alpha}\right)  ^{\alpha_{1}},\qquad
Y=Y_{0}\left(  t+\frac{\beta}{\alpha}\right)  ^{\alpha_{2}},\qquad
Z=Z_{0}\left(  t+\frac{\beta}{\alpha}\right)  ^{\alpha_{3}},
\end{equation}
with $X_{0},Y_{0},Z_{0}$ are integrating constants and $\left(  \alpha
_{i}\right)  _{i=1}^{3}\in\mathbb{R}.$ We may also check that the KSS vector
field must satisfies the relationship
\begin{equation}
\left[  KSS,\xi_{i}\right]  =C_{ij}^{k}\xi_{k},
\end{equation}
where $\xi$ is a Killing vector field, in this case $\xi_{i}=\partial_{i}. $
This relationship implies that the following equations must be satisfy:%
\begin{equation}
\delta+H_{i}\left(  \alpha t+\beta\right)  =0,
\end{equation}
where $H_{i}=\frac{X^{\prime}}{X}$ respectively, $i=1,2,3.$

In this way we find that%
\begin{equation}
H=\left(  \alpha_{1}+\alpha_{2}+\alpha_{2}\right)  \left(  t+\frac{\beta
}{\alpha}\right)  ^{-1}=A\left(  t+\gamma\right)  ^{-1},\qquad q=\frac{d}%
{dt}\left(  \frac{1}{H}\right)  -1=\frac{1}{A}-1,
\end{equation}
where $A=\left(  \sum_{i=1}^{3}\alpha_{i}\right)  ,\quad\frac{\beta}{\alpha
}=\gamma.$

Therefore, we may calculate the energy density from the conservation equation
i.e.%
\begin{equation}
\rho=\rho_{0}\left(  t+\gamma\right)  ^{-\left(  \omega+1\right)  A},
\end{equation}
so, following the same procedure as in the above sections, we find that%
\begin{equation}
\Lambda^{\prime}=-\frac{\tilde{A}}{c^{2}}\frac{2}{\left(  t+\gamma\right)
^{3}}-\frac{8\pi\rho_{0}}{c^{4}}\left(  G^{\prime}\left(  t+\gamma\right)
^{-(\omega+1)A}-(\omega+1)AG\left(  t+\gamma\right)  ^{-(\omega+1)A-1}\right)
,\label{laura2}%
\end{equation}
with $\tilde{A}=\alpha_{1}\alpha_{2}+\alpha_{3}\alpha_{1}+\alpha_{2}\alpha
_{3},$ so constant $G$ has the following behavior%
\begin{equation}
G=\frac{c^{2}}{8\pi\rho_{0}}\frac{2\tilde{A}}{(\omega+1)A}\left(
t+\gamma\right)  ^{(\omega+1)A-2},\label{laura5}%
\end{equation}
as we can see it is verified the relationship $G\rho\thickapprox\left(
t+\gamma\right)  ^{-2},$ as it is expected.

The behavior of $G$ is the following one:%
\begin{equation}
G=G_{0}\left(  t+\gamma\right)  ^{(\omega+1)A-2}\qquad\Longrightarrow
\qquad\left\{
\begin{array}
[c]{l}%
\text{growing if }(\omega+1)A>2\\
\text{constant if }(\omega+1)A=2\\
\text{decreasing if }(\omega+1)A<2
\end{array}
\right.  .
\end{equation}
with $G_{0}>0.$

Now, we next to calculate the quantity $\Lambda,$ therefore from eq.
(\ref{laura2}) it is found that%
\begin{equation}
\Lambda=\Lambda_{0}\left(  t+\gamma\right)  ^{-2},\qquad\Lambda_{0}%
=\frac{\tilde{A}}{c^{2}}\left(  1-\frac{2}{(\omega+1)A}\right)
,\label{laura6}%
\end{equation}
in this way it is observed that%
\begin{equation}
\Lambda=\Lambda_{0}\left(  t+\gamma\right)  ^{-2},\qquad\left\{
\begin{array}
[c]{c}%
\Lambda_{0}>0\Longleftrightarrow2<(\omega+1)A\\
\Lambda_{0}=0\Longleftrightarrow2=(\omega+1)A\\
\Lambda_{0}<0\Longleftrightarrow2>(\omega+1)A
\end{array}
\right.  .
\end{equation}

The shear behaves as
\begin{equation}
\sigma^{2}=\frac{1}{3c^{2}}\left(  \sum_{i}^{3}\alpha_{i}^{2}-\sum_{i\neq
j}^{3}\alpha_{i}\alpha_{j}\right)  \frac{1}{\left(  t+\gamma\right)  ^{2}}%
\neq0.
\end{equation}

As in the previous case, the SS solution, we suggest a way to calculate the
coefficients $\left(  \alpha_{i}\right)  $. They must to satisfy the field
equations i.e. they have to be solution of the following system of equations
\begin{align}
\alpha_{2}\left(  \alpha_{2}-1\right)  +\alpha_{3}\left(  \alpha_{3}-1\right)
+\alpha_{3}\alpha_{2}  &  =\tilde{A}\left(  \frac{A-2}{A}\right)
,\label{l1}\\
\alpha_{1}\left(  \alpha_{1}-1\right)  +\alpha_{3}\left(  \alpha_{3}-1\right)
+\alpha_{3}\alpha_{1}  &  =\tilde{A}\left(  \frac{A-2}{A}\right)
,\label{l2}\\
\alpha_{2}\left(  \alpha_{2}-1\right)  +\alpha_{1}\left(  \alpha_{1}-1\right)
+\alpha_{1}\alpha_{2}  &  =\tilde{A}\left(  \frac{A-2}{A}\right)  ,\label{l3}%
\end{align}
where $\tilde{A}=\alpha_{1}\alpha_{2}+\alpha_{3}\alpha_{1}+\alpha_{2}%
\alpha_{3},$ and $A=\alpha_{1}+\alpha_{2}+\alpha_{3}.$

So we have the following solutions for this system of equations:%
\begin{align}
\alpha_{1} &  =\alpha_{2}=\alpha_{3},\label{ayla1}\\
\alpha_{1} &  =1-\alpha_{2}-\alpha_{3},\label{ayla3}%
\end{align}
as it is observed we have found the same behavior as in the SS case, but in
this case the solution is nonsingular.

We would like to stress that a similar solution is already known since 1946 by
Narlikar and Karmarkar (see \cite{NK}). They found the following solution%
\begin{equation}
ds^{2}=dt^{2}-(kt+1)^{p}dx^{2}-(kt+1)^{q}dy^{2}-(kt+1)^{r}dz^{2},
\end{equation}
where $\left(  p,q,r\right)  $ must satisfy the following relationships:
$p+q+r=2,~pq+qr+rp=0.$

Before end we would like to make a little comment about the Kasner like
solutions. If a solution of (\ref{l1}-\ref{l3}) verifies the relationships
\begin{equation}
\sum_{i}^{3}\alpha_{i}^{2}=\sum_{i}^{3}\alpha_{i}=1,
\end{equation}
i.e. they are Kasner's type (see \cite{Kasner}, \cite{SH} and in particular
\cite{griego}), then this means that $\ \tilde{A}=\alpha_{1}\alpha_{2}%
+\alpha_{3}\alpha_{1}+\alpha_{2}\alpha_{3}=0,$ which brings us to get the
following result%
\begin{align}
\alpha_{1} &  =\frac{1}{2}\left(  1-\alpha_{3}-\sqrt{1+2\alpha_{3}-3\alpha
_{3}^{2}}\right)  <0,\forall\alpha_{3}\in\left(  0,1\right)  ,\label{maite1}\\
\alpha_{2} &  =\frac{1}{2}\left(  1-\alpha_{3}+\sqrt{1+2\alpha_{3}-3\alpha
_{3}^{2}}\right)  >0,\forall\alpha_{3}\in\left(  0,1\right)  ,\label{maite2}\\
\alpha_{3} &  =\alpha_{3},\label{maite3}%
\end{align}
we think that this class of solutions are unphysical and have a pathological
curvature behavior as it is shown bellow.

We will show that for this class of solutions we have the following curvature
behavior%
\begin{equation}
I_{1}=\frac{16\alpha^{4}\alpha_{3}^{2}\left(  1-\alpha_{3}\right)  }%
{c^{4}\left(  \alpha t+\beta\right)  ^{4}},\qquad I_{2}=0.
\end{equation}

The non-zero components of the Weyl tensor are:%
\begin{align}
C_{1212} &  =-\alpha_{2}\alpha_{3}\alpha^{2}\left(  \alpha t+\beta\right)
^{-\left(  1+\alpha_{3}+\Delta\right)  },\qquad C_{1313}=-\alpha_{1}\alpha
_{3}\alpha^{2}\left(  \alpha t+\beta\right)  ^{-\left(  1+\alpha_{3}%
+\Delta\right)  },\qquad C_{1414}=\alpha_{3}\alpha^{2}\left(  \alpha
t+\beta\right)  ^{-2\left(  1-\alpha_{3}\right)  }\nonumber\\
C_{2323} &  =\frac{\left(  \alpha_{3}-1\right)  \alpha_{3}\alpha^{2}}{c^{2}%
}\left(  \alpha t+\beta\right)  ^{-2\alpha_{3}},\qquad C_{2424}=\frac
{\alpha_{1}\alpha_{3}\alpha^{2}}{c^{2}}\left(  \alpha t+\beta\right)
^{-\left(  1-\alpha_{3}+\Delta\right)  },\qquad C_{3434}=\frac{\alpha
_{2}\alpha_{3}\alpha^{2}}{c^{2}}\left(  \alpha t+\beta\right)  ^{\left(
\alpha_{3}+\Delta-1\right)  },
\end{align}
where $\Delta=\sqrt{1+2\alpha_{3}-3\alpha_{3}^{2}}.$ The non-zero components
of the electric part of the Weyl tensor are:%
\begin{equation}
E_{22}=\frac{\alpha_{2}\alpha_{3}\alpha^{2}}{c^{2}}\left(  \alpha
t+\beta\right)  ^{-\left(  \alpha_{3}+\Delta+1\right)  },\qquad E_{33}%
=\frac{\alpha_{1}\alpha_{3}\alpha^{2}}{c^{2}}\left(  \alpha t+\beta\right)
^{-\left(  1+\alpha_{3}-\Delta\right)  },\qquad E_{44}=\frac{\left(
1-\alpha_{3}\right)  \alpha_{3}\alpha^{2}}{c^{2}}\left(  \alpha t+\beta
\right)  ^{-2\left(  1-\alpha_{3}\right)  },
\end{equation}
and the last invariant has the following behavior%
\begin{equation}
I_{3}=\frac{16\alpha^{4}\alpha_{3}^{2}\left(  1-\alpha_{3}\right)  }%
{c^{4}\left(  \alpha t+\beta\right)  ^{4}},
\end{equation}
while the gravitational entropy behaves as%
\begin{equation}
P^{2}=\infty,
\end{equation}
since $I_{2}=0.$ Note that if we take into account another definition for
$P^{2}=I_{3}/I_{1},$ \ (see \cite{gron1}-\cite{gron2}) then we get $P^{2}%
\neq\infty,$ since $I_{1}\neq0.$

Furthermore, as we can see, if $\tilde{A}=0,$ then from eqs. (\ref{laura5} and
\ref{laura6}) we get%
\begin{equation}
G=0,\qquad\Lambda=0,
\end{equation}
as it is expected for this class of solutions (vacuum solutions) so they are
not interested for us. Nevertheless relaxing the condition $\sum\alpha_{i}%
^{2}=1,$ to our result i.e. $\sum\alpha_{i}^{2}<1$, we are able to obtain
solutions whit $\left(  \alpha_{i}\right)  >0,\forall i$, and $G\neq
0,\Lambda\neq0.$

\section{Conclusions.\label{Conclu}}

We have shown how to attack a perfect fluid Bianchi I with $G$ and $\Lambda$
variable under the condition $\operatorname{div}T=0.$ In the appendix B we
will show how to modify these tactics to study this class of models but
relaxing the condition $\operatorname{div}T=0,$ i.e. without considering the
condition $\operatorname{div}T=0$ and taking into account only the hypothesis
of SS.

With the first of the exposed tactics, i.e. the Lie group one, we have solved
the field equations, solving only one ODE, eq. (\ref{defeq}), studying the
possible forms that takes $G(t)$ in order to make \ eq. (\ref{defeq})
integrable. We have started imposing a particular symmetry, $X=(at+e)\partial
_{t}+b\partial\rho$, to study all the possible symmetries would result a very
tedious work.

In this way we have obtained three exact solutions (well actually only two) in
function of the behavior of $G(t).$ We have seen that the scaling symmetry,
$X=at\partial_{t}+b\partial\rho$, brings us to get the already known flat FRW
solution since we have arrive to the conclusion that $\sigma=0$, i.e. the
shear vanish, and therefore we have rejected this solution since we are only
interested in solutions that verify the condition $\sigma\neq0.$ The second of
the obtained solutions, exponential behavior for $G(t) $, brings us to rule it
out, since this solution is only possible if $\rho=0.$ The last of the
obtained solutions, the third one, which is quite similar to the obtained one
by Kalligas et al (see \cite{We}), also has been ruled it out in spite of
seeming with physical meaning, $\sigma\neq0,$ etc...., since when we calculate
the numerical values of the exponents of the scale factors $\left(  \alpha
_{i}\right)  _{i=1}^{3},$ we shown that the only possible solution is the flat
FRW one but, which is more incredible, with $G=const.$ and $\Lambda$
vanishing. This has been a really surprising result, since we think that the
followed tactic, i.e. solving eq. (\ref{defeq}) without imposing any
assumption ad hoc, brings us to get consistent results, i.e. $\sigma\neq0$,
$G(t)\neq const.$ and of course, $\Lambda\neq0.$

It is clear that the latter solution is quite similar to the obtained one by
Kalligas et al, in fact, we have followed their method in order to arrive to
the same equation, but all the time, we have tried to avoid to assume ad hoc
any particular behavior for any of the quantities. Instead of following this
way, we have preferred to deduce from a symmetry principle the possible
behaviors for the function $G.$ Nevertheless Kalligas et at never arrive until
the last consequences in their calculations, since they did not try to find
the possible values for the exponents for the scale factors $\left(
\alpha_{i}\right)  _{i=1}^{3}$, for this reason their result looks with
physical meaning.

In appendix A we will show by solving the second order differential equation
(\ref{New-Lie}), using the same procedure as the exposed one in section IV,
that it is obtained, at least in order of magnitude, the same behavior as the
obtained one in section IV. Nevertheless, since we have two constrains, which
means that we have two integration constants, and therefore in principle they
are unknown we are introducing more uncertain in our solutions. Remember that
in section IV we had only one constrain, $G_{0},$ and this fact allows us to
arrive to a complete solution for each of the quantities. In this way, and
following the same procedure, in this case we are not able to rule the
solutions out, except in the case of exponential behavior, since we get
solutions with $\sigma\neq0$, although as we already know this is only a
mathematical drawback.

Nevertheless and knowing that this method has this kind of drawbacks we have
preferred to show both methods in other to show that at least in order of
magnitude, both methods arrive to the same conclusion i.e. we get the same
behavior. \ In a forthcoming paper we study a more complicated model (from the
mathematical point of view) which is a Bianchi I within the framework of
variable speed of light (VSL). In this case will be more useful to study the
second order ODE instead of the resulting third order ODE, which is really
complicate of studying from the point of view of the Lie method since this
equation will have four unknowns. Therefore, as we already know, we get the
same order of magnitude studying the second order ODE instead of the third
order ODE (other question will be how to solve the problem of the integrating
constants, but this will be other history).

At the same time we have shown that it is not necessary to make any ad hoc
assumption or to take into account any previous hypothesis or considering any
hypothetical behavior for any quantity since all these hypotheses could be
deduced from the symmetry principles, as for example using the Lie group
methods or studying the model from the point of view of the geometrical
symmetries i.e. SS etc...

With regard to the other tactics employed to study the field equations, i.e.
SS, MC and KSS, we have shown that both tactics are quite similar and that
they bring us to get really similar results, actually as we already know, with
the SS and the MC we get the same results.

We have shown that the solution obtained with the SS hypothesis is also quite
similar to the obtained one using the Lie method under the scale symmetry,
except than here, we get the important result $\sigma\neq0.$ This solution
also is valid for all equation of state i.e. $\forall\omega\in\left(
-1,1\right)  $, which enlarge the possibilities for this kind of solutions as
we will see in appendix B. Furthermore, the exponents must satisfy the
following relationships $\sum\alpha_{i}=1,$ and $\sum\alpha_{i}^{2}<1.$ We
would like to point out that if $\omega=1,$ then we regain the classical
solution where $G$ behaves as a true constant while $\Lambda$ vanish. In
appendix B we have tried to show that this class of solutions are consistent
with previous results obtained by other authors. Nevertheless other authors
that have studied this kind of models (under the SS as well as KSS hypothesis,
but with constants as true constants and therefore with $\Lambda$ vanishing)
have arrived to the conclusion that these solutions must satisfy the
relationships $\sum\alpha_{i}=1,$ and $\sum\alpha_{i}^{2}=1$. We think that
this class of solutions are unphysical since necessarily one of the scale
factors must be a decreasing time function (maybe such class of solutions
would have any interest in the study of singularities). Furthermore such
conditions are quite restrictive and impose strong restrictions to the
curvature tensors, for example, the model is Ricci flat which means that
$I_{2}=0,$ so the gravitational entropy is infinite.

With regard to the behavior of the \textquotedblleft
constants\textquotedblright\ $G$ and $\Lambda$ we would like to stress that we
have arrive to some surprising results since $\forall\omega\in\left(
-1,1\right)  $, $G$ is a decreasing function on time while $\Lambda$ is also a
decreasing time function but negative i.e. $\Lambda<0.$ In the same framework,
for the case of the flat FRW, always $G$ is growing. The main difference
between the SS and the KSS solution is that the KSS one is nonsingular while
the SS one is singular.

We furthermore have pointed out, as it is well known, that if the ST is SS
then there is a vector field, $V\in\mathfrak{X}(M)$ that satisfies the
equation $\mathcal{L}_{V}g=2g,$ then such vector field must satisfy the
equation $\mathcal{L}_{V}T=0,$ i.e. a homothetic vector field is also a MC
vector field. Modifying in an appropriate way the MC equations we have been
able to find the same relationships as in the case of the SS solution.
Therefore we have shown that this tactic would be very useful in the study of
more complicated models as for example the viscous ones.

To end, we would like to comment the obtained results in appendix B. In this
appendix we have found four exact solutions for different Bianchi I models
under the SS hypothesis. In the first of them, we study the standard Bianchi I
model i.e. which where $G=const.$ and $\Lambda=0.$ We have found again the
solution already obtained for many authors, but with the restrictions
$\sum\alpha_{i}=1,$ $\sum\alpha_{i}^{2}<1$ iff $\omega=1.$

In the other studied cases we consider the possibility of one of the constants
vary as well as both vary at the same time but with the condition
$\operatorname{div}T\neq0.$ Therefore, in the model with only $G$ time-varying
\ we find again, as in the previous case, that it is only possible if are
verified the same conditions i.e. $\sum\alpha_{i}=1,$ $\sum\alpha_{i}^{2}<1$
iff $\omega=1.$ If $\omega\neq1,$ then the models collapses to the flat FRW
one. Nevertheless in the case where $\Lambda$ vary or both constants vary we
show \ that such possibilities are possible iff $\sum\alpha_{i}=1,$
$\sum\alpha_{i}^{2}<1$ and $\omega\neq1.$ We find that $G $ is a decreasing
time function on time while $\Lambda$ is a negative decreasing time function.
We have shown how to regain, in a trivial way, the condition
$\operatorname{div}T=0.$

\begin{acknowledgments}
I would like to thank to G.S. Hall and T. Harko their comments and specially
to Laura Fern\'{a}ndez for her support.
\end{acknowledgments}

\appendix

\section{Lie Method.\label{ApA}}

As we said in the introduction, we would like to compare some different
techniques in order to be sure which of them is better when one is studying
more complicate models, as for example a Bianchi I with c-var. For this
purpose, in this section, we study the set of solutions for the eq.%
\begin{equation}
\ddot{\rho}=K_{1}\frac{\dot{\rho}^{2}}{\rho}+K_{2}G\rho^{2}-K_{3}\Lambda
\rho,\label{ayasa1}%
\end{equation}
where%
\begin{equation}
K_{1}=\frac{2+\omega}{1+\omega},\qquad K_{2}=\frac{12\pi\left(  \omega
^{2}-1\right)  }{c^{2}},\qquad K_{3}=3\left(  1+\omega\right)  c^{2}.
\end{equation}
note that we have the very simple case $K_{2}=0\Longleftrightarrow\omega=1,$
i.e.
\begin{equation}
\ddot{\rho}=\frac{3}{2}\frac{\dot{\rho}^{2}}{\rho}-6c^{2}\Lambda
\rho.\label{ayasa2}%
\end{equation}

Observe that we have a second order ODE with three unknowns, which looks
simpler than the previous case (see section \ref{LM}), but actually, as we
will show, this tactic has many drawbacks, for example the obtained solutions
depend on many integrating constants and hence it is quite difficult to
\textquotedblleft predict\textquotedblright\ their behavior or to get rid of
some of them.

Following the previous procedure we get:%

\begin{align}
K_{1}\xi_{\rho}+\rho\xi_{\rho\rho} &  =0,\\
K_{1}\left(  \eta-\rho\eta_{\rho}\right)  +\rho^{2}\left(  \eta_{\rho\rho
}-2\xi_{t\rho}\right)   &  =0,\\
3\rho^{3}\xi_{\rho}\left(  K_{3}\Lambda-K_{2}\rho G\right)  -2K_{1}\rho
\eta_{t}+2\rho^{2}\eta_{t\rho}-\xi_{tt}\rho^{2} &  =0,\\
\eta\rho^{2}\left(  K_{3}\Lambda-K_{2}\rho G\right)  +2\rho^{3}\xi_{t}\left(
K_{3}\Lambda-K_{2}\rho G\right)  -\rho^{3}\eta_{\rho}\left(  K_{3}%
\Lambda-K_{2}\rho G\right)  +\xi\rho^{3}\left(  K_{3}\Lambda^{\prime}%
-K_{2}\rho G^{\prime}\right)  +\rho^{2}\eta_{tt} &  =0,\label{auro1}%
\end{align}
where from (\ref{auro1}) we get the following constrains if we impose the
scaling symmetry $X=\left(  at+e\right)  \partial_{t}+b\rho\partial_{\rho}, $
(as in the above case)%
\begin{align}
\frac{\Lambda^{\prime}}{\Lambda} &  =-\frac{2a}{at+e},\label{restric1}\\
\frac{G^{\prime}}{G} &  =-\frac{b+2a}{at+e},\label{restric2}%
\end{align}

Therefore we go next to study the set of the possible solutions. As in section
\ref{LM}, we may follow the standard procedure but we would like to point out
(to stress) that in this case we have two integration constant $\Lambda_{0}$
and $G_{0}$ instead of only one as above so we are introducing more uncertain
in this approach. Therefore we find again three solutions, the scaling and the
full symmetry and the exponential one. The scaling solution has the same
behavior as the obtained one in section (\ref{LM}), at least in order of
magnitude, but we are not able to rule it out since depends of a lot of
numerical constants so for example we do not arrive to the conclusion that
$\sigma=0,$ as above. With the regard to the full symmetry we have the same
history, we find that this solution has the same order of magnitude and that
it is nonsingular but depends of many integration as well as numerical
constants so the obtained solution is very imprecise although it looks with
physical meaning. To end and as in the above solution we are also able to get
rid of the exponential solution, in this case $\Lambda=\Lambda_{0}=const.$ and
therefore $\rho_{0}=0.$ We may see, for example, how works all this procedure
in the case of the scaling symmetry because the rest of the cases are exactly
the same as the exposed ones in section \ref{LM}.

\subsection{Scaling symmetry}

In this case we have $X=at\partial_{t}+b\rho\partial_{\rho}$ and hence%
\begin{align}
\frac{\Lambda^{\prime}}{\Lambda} &  =-\frac{2}{t}\qquad\Longrightarrow
\qquad\Lambda=\Lambda_{0}t^{-2},\qquad\Lambda_{0}\in\mathbb{R},\label{monic1}%
\\
\frac{G^{\prime}}{G} &  =-\frac{b+2a}{at}\qquad\Longrightarrow\qquad
G=G_{0}t^{-2-\frac{b}{a}},\qquad G_{0}\in\mathbb{R}^{+},\label{monic2}%
\end{align}
i.e. we are \textquotedblleft assuming\textquotedblright\ that $\Lambda_{0}$
is a real constant (but at this point, we do not know which is its sign) and
$G_{0}$ is a positive real constant, while%
\begin{equation}
\frac{dt}{at}=\frac{d\rho}{b\rho}\qquad\Longrightarrow\qquad\rho=\rho
_{0}t^{\frac{b}{a}},\qquad\rho_{0}\in\mathbb{R}^{+},
\end{equation}
making the assumption, $ab<0$ with $b<0,$ and where $\rho_{0}$ is a positive
real constant. As we can see, it is verified the relationship $G\rho=t^{-2},$
i.e. the Mach relationship for the inertia.

Therefore%
\begin{equation}
\rho\thickapprox t^{\frac{b}{a}}\text{ is decreasing,\qquad}G\thickapprox
t^{-2-\frac{b}{a}}\left\{
\begin{array}
[c]{l}%
\text{growing if }-b>2a\\
\text{constant if }-b=2a\\
\text{decreasing if }-b<2a
\end{array}
\right.  ,\qquad\Lambda=\Lambda_{0}t^{-2},
\end{equation}
but at this point we have not any information about the sing of $\Lambda_{0}.$

Since $\rho$ must verifies eq. (\ref{monic1}) with $G(t)$ and \ $\Lambda(t) $
given by eqs. (\ref{monic1}-\ref{monic2}), we find $\rho_{0},$
\begin{equation}
\rho_{0}=\frac{c^{2}\left(  ab\left(  1+\omega\right)  +b^{2}-3c^{2}%
\Lambda_{0}a^{2}\left(  1+\omega\right)  ^{2}\right)  }{G_{0}12\pi
a^{2}\left(  1+\omega\right)  ^{2}\left(  1-\omega\right)  },
\end{equation}
with the only assumption $\omega\in\left(  -1,1\right)  ,$i.e. $\omega$
$\neq-1,$ and $\omega\neq1$%
\begin{equation}
\rho_{0}>0\Longleftrightarrow b^{2}>-ab\left(  1+\omega\right)  +3c^{2}%
\Lambda_{0}a^{2}\left(  1+\omega\right)  ^{2},
\end{equation}
with $ab<0.$ Furthermore, $\rho$ must verify eq. (\ref{eq9}) so
\begin{equation}
\Lambda^{\prime}=-\frac{8\pi G^{\prime}}{c^{4}}\rho\Longrightarrow-\Lambda
_{0}=\left(  2+\frac{b}{a}\right)  \frac{4\pi}{c^{4}}G_{0}\rho_{0}%
\Longrightarrow\rho_{0}=-\frac{a\Lambda_{0}c^{4}}{4\pi G_{0}\left(
2a+b\right)  },
\end{equation}
and hence%
\begin{equation}
c^{2}\Lambda_{0}=-\frac{b}{3a\left(  1+\omega\right)  }\left(  1+\frac
{b}{3a\left(  1+\omega\right)  }\right)  \left(  \frac{2a+b}{a\left(
1-\omega\right)  +2a+b}\right)  .\label{desco}%
\end{equation}
$\allowbreak$

\begin{remark}
As it is observed we are obtaining the same order of magnitude for each
quantity, but in this case, we have less information about the behavior of the
numerical constants, since they depend on more integrating constants.
\end{remark}

As in the above section, with regard to $H$ we find that
\begin{equation}
Kt^{\alpha}=R_{0}t^{-b/a\left(  1+\omega\right)  },\qquad\Longrightarrow
\qquad\sum_{i}^{3}\alpha_{i}=\alpha=-\frac{b}{^{a\left(  1+\omega\right)  }},
\end{equation}
as in the previous cases.

The shear has the following behavior$.$%
\begin{equation}
\sigma^{2}=\frac{1}{3c^{2}\left(  1+\omega\right)  ^{2}}\left(  \frac
{\dot{\rho}}{\rho}\right)  ^{2}-\left(  8\pi\frac{G}{c^{4}}\rho+\Lambda
\right)  =\left(  \frac{\omega+1}{1-\omega}\right)  \left[  \Lambda_{0}%
-\frac{b^{2}}{a^{2}}\frac{1}{3c^{2}\left(  1+\omega\right)  ^{2}}\left(
1+2\frac{a}{b}\right)  \right]  t^{-2}=\sigma_{0}^{2}t^{-2},
\end{equation}
$\allowbreak$therefore we have the following possibilities $\forall\omega
\in\left(  -1,1\right)  .$ Note that if $\omega<-1$ (phantom case) then we
need to make other considerations. Since $\sigma^{2}>0,$ note that $\frac
{a}{b}<0,$ then we may suppose that $\Lambda_{0}>0$ and then%
\begin{align}
b+2a &  <0\qquad\Longrightarrow\qquad\Lambda_{0}-\frac{b^{2}}{a^{2}}\frac
{1}{3c^{2}\left(  1+\omega\right)  ^{2}}\left(  1+2\frac{a}{b}\right)  >0,\\
b+2a &  =0\qquad\Longrightarrow\qquad\sigma_{0}^{2}=0,\text{ \ for
eq.(\ref{desco}),}\\
b+2a &  >0\qquad\Longrightarrow\qquad\Lambda_{0}>\left\vert \frac{b^{2}}%
{a^{2}}\frac{1}{3c^{2}\left(  1+\omega\right)  ^{2}}\left(  1+2\frac{a}%
{b}\right)  \right\vert ,
\end{align}
but if $\Lambda_{0}<0$ then we have the next possibilities%
\begin{equation}
\frac{b^{2}}{a^{2}}\frac{1}{3c^{2}\left(  1+\omega\right)  ^{2}}\left(
1+2\frac{a}{b}\right)  >\left\vert \Lambda_{0}\right\vert ,
\end{equation}
with $b+2a<0,$ otherwise the solutions lack of any mathematical meaning and
hence of physical one. In this way we note that we have loss of information
since we have two integration constant, $G_{0}$ and $\Lambda_{0}$ while in the
previous case we were able to determine perfectly the behavior of each
quantity and in this case we have less information about their behavior
although both cases are quite similar at least in order of magnitude.

\begin{remark}
Therefore, we may say that this tactic is also valid, but has the strong
drawback of giving us less information about the obtained solution.
\end{remark}


\section{On self-similar solutions.\label{ApB}}

In this appendix we are going to study several Bianchi I models and we will
show how it is possible to find exact solutions to the field equations
(without the condition $divT=0)$ under the hypothesis of SS.

The time derivatives of $G$ and $\Lambda$ are related by the Bianchi
identities i.e. eq. (\ref{eq8}) that in this case collapses to the following
one:%
\begin{equation}
\dot{\rho}+\rho\left(  1+\omega\right)  \left(  \frac{\dot{X}}{X}+\frac
{\dot{Y}}{Y}+\frac{\dot{Z}}{Z}\right)  =f(t),\label{B1}%
\end{equation}
where $f(t)$ is a function that depends on time and controls the time
variation of the constant $G$ or/and $\Lambda.$ If $G=const.$ and $\Lambda$
vanish then $f(t)=0,$ so the model collapses to the standard one. This idea
was pointed out by Rastal (see \cite{Rastall}) and improved (in the
theoretical framework of time varying constants) by Harko and Mak (see
\cite{harko}).

Therefore the resulting field equations are (\ref{eq3}-\ref{eq6}) together to
the new one
\begin{equation}
\dot{\rho}+\rho\left(  1+\omega\right)  H=f(t),\label{Beq9}%
\end{equation}
with
\begin{equation}
H=\left(  \frac{\dot{X}}{X}+\frac{\dot{Y}}{Y}+\frac{\dot{Z}}{Z}\right)
=3\frac{\dot{R}}{R}\text{ \ and \ \ }R^{3}=XYZ,
\end{equation}
i.e. we are following the same notation of the previous sections.

We remember that the homothetic vector field is:%
\begin{equation}
V=t\partial_{t}+\left(  1-t\frac{\dot{X}}{X}\right)  x\partial_{x}+\left(
1-t\frac{\dot{Y}}{Y}\right)  y\partial_{y}+\left(  1-t\frac{\dot{Z}}%
{Z}\right)  z\partial_{z},\label{BHO1}%
\end{equation}
so this means that we have obtained the following behavior for the scale
factors:%
\begin{equation}
X=X_{0}t^{\alpha_{1}},\qquad Y=Y_{0}t^{\alpha_{2}},\qquad Z=Z_{0}t^{\alpha
_{3}},
\end{equation}
with $X_{0},Y_{0},Z_{0}$ are integrating constants and $\left(  \alpha
_{i}\right)  _{i=1}^{3}\in\mathbb{R}.$ In this way we find that%
\begin{equation}
H=\left(  \sum_{i=1}^{3}\alpha_{i}\right)  \frac{1}{t}=\frac{\alpha}{t},\qquad
q=\frac{d}{dt}\left(  \frac{1}{H}\right)  -1=\frac{1}{\alpha}-1,\qquad
\sigma^{2}=\frac{1}{3c^{2}}\left(  \sum_{i}^{3}\alpha_{i}^{2}-\sum_{i\neq
j}^{3}\alpha_{i}\alpha_{j}\right)  \frac{1}{t^{2}}.
\end{equation}

\subsection{\textquotedblleft Constants\textquotedblright\ constant.}

In this case we consider $f(t)=0,$ so this means that $G=const.$ and $\Lambda$
vanish and therefore we get that from eq. (\ref{Beq9}) that
\begin{equation}
\dot{\rho}+\rho\left(  1+\omega\right)  H=0,\qquad\Rightarrow\qquad\rho
=\rho_{0}t^{-(\omega+1)\alpha}.
\end{equation}

From the field equations (\ref{eq3}) we get that%
\begin{equation}
\rho_{0}=\frac{Ac^{2}}{8\pi G},\qquad\alpha=\frac{2}{(\omega+1)},
\end{equation}
where $A=\alpha_{1}\alpha_{2}+\alpha_{3}\alpha_{1}+\alpha_{2}\alpha_{3}.$

The shear has the following behavior, $\sigma^{2}\neq0$, as it is observed
$\sigma\rightarrow0$ as $\left(  \alpha_{i}\rightarrow\alpha_{j}\right)  . $
As in the previous sections, we may calculate the coefficients $\left(
\alpha_{i}\right)  $ by solving the following system of equations:%
\begin{align}
\alpha_{2}\left(  \alpha_{2}-1\right)  +\alpha_{3}\left(  \alpha_{3}-1\right)
+\alpha_{3}\alpha_{2} &  =-A\omega,\\
\alpha_{1}\left(  \alpha_{1}-1\right)  +\alpha_{3}\left(  \alpha_{3}-1\right)
+\alpha_{3}\alpha_{1} &  =-A\omega,\\
\alpha_{2}\left(  \alpha_{2}-1\right)  +\alpha_{1}\left(  \alpha_{1}-1\right)
+\alpha_{1}\alpha_{2} &  =-A\omega,\\
\alpha(\omega+1) &  =2,
\end{align}
where $A=\alpha_{1}\alpha_{2}+\alpha_{3}\alpha_{1}+\alpha_{2}\alpha_{3},$ and
$\alpha=\alpha_{1}+\alpha_{2}+\alpha_{3}.$

So we have the following solutions for this system of equations:%
\begin{align}
\alpha_{1} &  =1-\alpha_{2}-\alpha_{3},\qquad\omega=1,\label{Bsolss1}\\
\alpha_{1} &  =\alpha_{2}=\alpha_{3}=\frac{2}{3\left(  \omega+1\right)
},\label{Bsolss2}%
\end{align}
as it is observed only solution (\ref{Bsolss1}) is interesting for us. The
second solution is the usual FRW one, so it is not \ interesting for us (see
Einstein\&de Sitter (\cite{EdS}) for $\omega=0$, and Harrison (\cite{Harrison}%
) $\forall\omega)$. Nevertheless we have found that solution (\ref{Bsolss1})
verifies the conditions
\begin{equation}
\alpha=\sum\alpha_{i}=1,\qquad\sum\alpha_{i}^{2}<1,
\end{equation}
but iff $\omega=1.$ (see \cite{HW}) while other authors claim that must be
satisfies the condition $\sum\alpha_{i}^{2}=1,$ (see \cite{Kasner}, \cite{SH}
and \cite{Jacobs}) and in particular, in this context (see \cite{griego}).

Therefore we have obtained the following behavior for the main quantities:%
\begin{equation}
H=\frac{1}{t},\qquad\Longrightarrow\qquad q=0,
\end{equation}
so it is quite difficult to reconcile this model with the observational data.
With regard to the energy density we find that%
\begin{equation}
\rho=\frac{Ac^{2}}{8\pi G}t^{-2},\qquad\sigma^{2}=\frac{1}{3c^{2}}\left(
1+3A\right)  \frac{1}{t^{2}},\label{results1}%
\end{equation}
and with regard to the constants $\left(  \alpha_{i}\right)  _{i=1}^{3}$ we
have that only obtain a BI solution iff $\alpha_{1}=1-\alpha_{2}-\alpha_{3},$
(where furthermore we suppose that $\alpha_{2}\neq\alpha_{3})$ and that this
result only is possible if the equation of state is $\omega=1,$ i.e.
ultra-stiff matter (see \cite{HW}). For a review of Bianchi I solutions see
for example (\cite{Harvey}).

\subsection{$G-$variable.}

In this case we are going to consider that only vary \textquotedblleft
constant\textquotedblright\ $G.$ This only possible if we take into account
the condition $divT\neq0$ and therefore $f(t)=-\frac{G^{\prime}}{G}\rho,$ so
eq. (\ref{Beq9}) collapses to the following one.
\begin{equation}
\frac{\rho^{\prime}}{\rho}+\frac{G^{\prime}}{G}=-\left(  1+\omega\right)
\frac{\alpha}{t},\qquad\Longrightarrow\qquad\rho G=t^{-\left(  1+\omega
\right)  \alpha},\label{Blau1}%
\end{equation}

From the field equations (\ref{eq3}) we get that%
\begin{equation}
G\rho=\frac{c^{2}}{4\pi}\frac{A}{\alpha\left(  1+\omega\right)  }\frac
{1}{t^{2}},\qquad\alpha=\frac{2}{(\omega+1)}.
\end{equation}

The shear has the following behavior, $\sigma^{2}\neq0$, as it is observed
$\sigma\rightarrow0$ as $\left(  \alpha_{i}\rightarrow\alpha_{j}\right)  . $
As in the previous sections, to calculate the coefficients $\left(  \alpha
_{i}\right)  $ we need to solve the following system of equations:%
\begin{align}
\alpha_{2}\left(  \alpha_{2}-1\right)  +\alpha_{3}\left(  \alpha_{3}-1\right)
+\alpha_{3}\alpha_{2} &  =-A\omega,\\
\alpha_{1}\left(  \alpha_{1}-1\right)  +\alpha_{3}\left(  \alpha_{3}-1\right)
+\alpha_{3}\alpha_{1} &  =-A\omega,\\
\alpha_{2}\left(  \alpha_{2}-1\right)  +\alpha_{1}\left(  \alpha_{1}-1\right)
+\alpha_{1}\alpha_{2} &  =-A\omega,\\
\alpha(\omega+1) &  =2,
\end{align}
where $A=\alpha_{1}\alpha_{2}+\alpha_{3}\alpha_{1}+\alpha_{2}\alpha_{3},$ and
$\alpha=\alpha_{1}+\alpha_{2}+\alpha_{3}.$ So we have the same solution as in
the above case.

Therefore we have obtained the following behavior for the main quantities:%
\begin{equation}
H=\frac{1}{t},\qquad\Longrightarrow\qquad q=0,
\end{equation}
and
\begin{equation}
G\rho=\frac{Ac^{2}}{8\pi}t^{-2},\qquad\sigma^{2}=\frac{1}{3c^{2}}\left(
1-3A\right)  \frac{1}{t^{2}},
\end{equation}
note that this result is quite similar to the obtained one in the last
solution i.e. the obtained one in eq. (\ref{results1}), but we are not able to
get a separate behavior for the quantities $G$ and $\rho$.

\subsection{$\Lambda-$variable.}

In this case we consider only the variation of the cosmological constant
$\Lambda,$ so eq. (\ref{Beq9}) yields
\begin{equation}
\dot{\rho}+\rho\left(  1+\omega\right)  \left(  \frac{\dot{X}}{X}+\frac
{\dot{Y}}{Y}+\frac{\dot{Z}}{Z}\right)  =-\frac{\dot{\Lambda}c^{4}}{8\pi
G},\label{laugL}%
\end{equation}
and therefore from the field equations (\ref{eq3}) we get that%
\begin{equation}
\Lambda^{\prime}=-\frac{A}{c^{2}}\frac{2}{t^{3}}-\frac{8\pi G}{c^{4}}%
\rho^{\prime},\label{dorota1}%
\end{equation}
and hence%
\begin{equation}
\rho=\frac{c^{2}}{4\pi G}\frac{A}{\left(  1+\omega\right)  \alpha}\frac
{1}{t^{2}}.
\end{equation}

Now, we next to calculate the quantity $\Lambda,$ from eq. (\ref{dorota1}) we
get%
\begin{equation}
\Lambda=\frac{A}{c^{2}}\left(  1-\frac{2}{\left(  1+\omega\right)  \alpha
}\right)  \frac{1}{t^{2}},
\end{equation}
in this way it is observed that%
\begin{equation}
\Lambda=\Lambda_{0}t^{-2},\qquad\left\{
\begin{array}
[c]{c}%
\Lambda_{0}>0\Longleftrightarrow(\omega+1)\alpha>2\\
\Lambda_{0}=0\Longleftrightarrow(\omega+1)\alpha=2\\
\Lambda_{0}<0\Longleftrightarrow(\omega+1)\alpha<2
\end{array}
\right.  .
\end{equation}

The shear has the following behavior, $\sigma^{2}\neq0$, by hypothesis. As in
the previous sections, we calculate the coefficients $\left(  \alpha
_{i}\right)  $ from following system of equations:%
\begin{align}
\alpha_{2}\left(  \alpha_{2}-1\right)  +\alpha_{3}\left(  \alpha_{3}-1\right)
+\alpha_{3}\alpha_{2} &  =A\left(  \frac{\alpha-2}{\alpha}\right)  ,\\
\alpha_{1}\left(  \alpha_{1}-1\right)  +\alpha_{3}\left(  \alpha_{3}-1\right)
+\alpha_{3}\alpha_{1} &  =A\left(  \frac{\alpha-2}{\alpha}\right)  ,\\
\alpha_{2}\left(  \alpha_{2}-1\right)  +\alpha_{1}\left(  \alpha_{1}-1\right)
+\alpha_{1}\alpha_{2} &  =A\left(  \frac{\alpha-2}{\alpha}\right)  ,
\end{align}
where $A=\alpha_{1}\alpha_{2}+\alpha_{3}\alpha_{1}+\alpha_{2}\alpha_{3},$ and
$\alpha=\alpha_{1}+\alpha_{2}+\alpha_{3}.$

So we have the following solutions for this system of equations:%
\begin{align}
\alpha_{1} &  =\alpha_{2}=\alpha_{3},\label{Bsolsys1}\\
\alpha_{1} &  =1-\alpha_{2}-\alpha_{3},\label{Bsolsys3}%
\end{align}
as it is observed solution (\ref{Bsolsys1}) is not interesting for us, since
it is unphysical (in this context). Only the second solution has physical
meaning and it is valid $\forall\omega\in\left(  -1,1\right]  $. Nevertheless
we have found that this solution only verifies the first of the condition of
the Kasner like solutions i.e.
\begin{equation}
\alpha=\sum\alpha_{i}=1,\qquad and\qquad\sum\alpha_{i}^{2}<1,
\end{equation}
and it is valid $\forall\omega\in\left(  -1,1\right]  .$

Therefore we have obtained the following behavior for the main quantities:%
\begin{equation}
H=\frac{1}{t},\qquad\Longrightarrow\qquad q=0,
\end{equation}
while with regard to the energy density we find that%
\begin{equation}
\rho=\frac{c^{2}}{4\pi G}\frac{A}{\left(  1+\omega\right)  }\frac{1}{t^{2}%
},\text{ }%
\end{equation}
\ so, if $\omega<-1\Longrightarrow\rho$ is negative (phantom cosmologies), for
the rest of the values of $\omega,$ i.e. $\omega\in(-1,1],$ $\rho$ is a
decreasing function on time.

The cosmological \textquotedblleft constant\textquotedblright\ behaves as
follows
\begin{equation}
\Lambda=\Lambda_{0}t^{-2},\qquad\Lambda_{0}=\left\{
\begin{array}
[c]{l}%
\Lambda_{0}=0\Longleftrightarrow\omega=1,\\
\Lambda_{0}<0,\qquad\forall\text{ }\omega\in(-1,1)
\end{array}
\right.  ,
\end{equation}
so we have found that $\Lambda$ is a \textquotedblleft\emph{negative
decreasing function}\textquotedblright\ on time. As we can see this solution
is quite different of the previous ones, since here we have obtained a
solution type Bianchi I $\forall$ $\omega\in(-1,1)$ while in the previous ones
this only happens if $\omega=1.$ Here if $\omega=1$ then we regain the first
of the studied cases i.e. which one where $\Lambda$ vanish and $G$ behaves as
a true constant.

\subsection{$G\&\Lambda-$variable.}

In this case we are going to consider that both \textquotedblleft
constants\textquotedblright\ $G$ and $\Lambda$ vary, as in section \ref{SS},
but in this occasion it is not verified the additional condition $divT=0,$
therefore eq. (\ref{Beq9}) yields%

\begin{equation}
\frac{8\pi G}{c^{4}}\left[  \dot{\rho}+\rho\left(  1+\omega\right)  \left(
\frac{\dot{X}}{X}+\frac{\dot{Y}}{Y}+\frac{\dot{Z}}{Z}\right)  \right]
=-\dot{\Lambda}-\frac{8\pi}{c^{4}}\dot{G}\rho,\label{dorota2}%
\end{equation}
so from the field equations (\ref{eq3}) and (\ref{eq9}) we get that%
\begin{equation}
-\frac{2A}{t^{3}}=-\frac{8\pi}{c^{2}}\left(  G^{\prime}\rho+\rho^{\prime
}G\right)  +\Lambda^{\prime}c^{2},
\end{equation}
and taking into account the eq.
\begin{equation}
\dot{\rho}+\rho\left(  1+\omega\right)  H=-\frac{\dot{\Lambda}c^{4}}{8\pi
G}-\frac{\dot{G}}{G}\rho,
\end{equation}
we get%
\begin{equation}
-\frac{2A}{t^{3}}=-\frac{8\pi}{c^{2}}\left(  -G\rho\left(  1+\omega\right)
\frac{\alpha}{t}\right)  ,
\end{equation}
and hence
\begin{equation}
G\rho=\frac{c^{2}}{4\pi}\frac{A}{(\omega+1)\alpha}t^{-2}.
\end{equation}
as we can see it is verified the relationship $G\rho\thickapprox t^{-2},$ as
it is expected. In fact it is impossible to separate both functions (to get
the behavior of both functions independently), to do that we need to impose a
condition, but precisely we are trying to avoid such way.

Now taking into account again eq. (\ref{eq3}) we get%
\begin{equation}
\Lambda=\Lambda_{0}t^{-2},\qquad\Lambda_{0}=\frac{A}{c^{2}}\left(  1-\frac
{2}{(\omega+1)\alpha}\right)  ,
\end{equation}
in this way it is observed that%
\begin{equation}
\Lambda=\Lambda_{0}t^{-2},\qquad\left\{
\begin{array}
[c]{c}%
\Lambda_{0}>0\Longleftrightarrow(\omega+1)\alpha>1\\
\Lambda_{0}=0\Longleftrightarrow(\omega+1)\alpha=1\\
\Lambda_{0}<0\Longleftrightarrow(\omega+1)\alpha<1
\end{array}
\right.  .
\end{equation}

The shear behaves (see eq. (\ref{defshear})) as follows: $\sigma^{2}\neq0,$ by
hypothesis. In order to find the value of constants $\left(  \alpha
_{i}\right)  ,$ we make that them verify the field eqs. so in this case we get
the following system of eqs.:%
\begin{align}
\alpha_{2}\left(  \alpha_{2}-1\right)  +\alpha_{3}\left(  \alpha_{3}-1\right)
+\alpha_{3}\alpha_{2} &  =A\left(  \frac{\alpha-2}{\alpha}\right)  ,\\
\alpha_{1}\left(  \alpha_{1}-1\right)  +\alpha_{3}\left(  \alpha_{3}-1\right)
+\alpha_{3}\alpha_{1} &  =A\left(  \frac{\alpha-2}{\alpha}\right)  ,\\
\alpha_{2}\left(  \alpha_{2}-1\right)  +\alpha_{1}\left(  \alpha_{1}-1\right)
+\alpha_{1}\alpha_{2} &  =A\left(  \frac{\alpha-2}{\alpha}\right)  ,
\end{align}
where $A=\alpha_{1}\alpha_{2}+\alpha_{3}\alpha_{1}+\alpha_{2}\alpha_{3},$ and
$\alpha=\alpha_{1}+\alpha_{2}+\alpha_{3}.$ Therefore we obtain the same
solution as in the last studied case finding in this way the following
behavior for the main quantities:%
\begin{equation}
H=\frac{1}{t},\qquad\Longrightarrow\qquad q=0,
\end{equation}
and \ with regard to the product $G\rho$ we get
\begin{equation}
G\rho=\frac{c^{2}}{4\pi}\frac{A}{(\omega+1)}t^{-2},
\end{equation}
but we cannot say anything more. The cosmological \textquotedblleft constant"
behaves as follows
\begin{equation}
\Lambda=\Lambda_{0}t^{-2},\qquad\Lambda_{0}=\left\{
\begin{array}
[c]{l}%
\Lambda_{0}=0\Longleftrightarrow\omega=1\\
\Lambda_{0}<0,\qquad\forall\text{ }\omega\in(-1,1)
\end{array}
\right.  ,
\end{equation}
so we have found that $\Lambda$ is a negative decreasing function on time.

In order to try to find a separate behavior for the functions $\rho$ and $G, $
we may suppose that
\begin{equation}
\rho=\rho_{0}t^{-a},\qquad G=G_{0}t^{a-2},\qquad\Longrightarrow\qquad
G\rho=\frac{c^{2}}{4\pi}\frac{A}{(\omega+1)}t^{-2}=Kt^{-2},
\end{equation}
with $a\in\mathbb{R}^{+},$ i.e. for example we may choice
\begin{equation}
G=\frac{c^{2}}{4\pi\rho_{0}}\frac{A}{(\omega+1)}t^{a-2}=\frac{K}{\rho_{0}%
}t^{a-2},
\end{equation}
therefore, it is verified the field eq. (\ref{dorota2}) for all the possible
values of $a,$ but if $a=(\omega+1)$ then we regain the condition
$\operatorname{div}T=0$ as well as $f(t)=0,$ i.e.
\begin{equation}
\operatorname{div}T=\dot{\rho}+\rho\left(  1+\omega\right)  H=0=-\frac
{\dot{\Lambda}c^{4}}{8\pi G}-\frac{\dot{G}}{G}\rho=f(t).
\end{equation}

\end{document}